\documentclass[a4paper,aps,twocolumn,groupedaddress]{revtex4} 
\usepackage{graphicx}  
\usepackage{dcolumn}   
\usepackage{bm}        
\usepackage{amssymb,amsmath,dsfont}   
\usepackage{color}

\begin{document}

\title{Instability-induced fermion production in quantum field theory}
\author{J{\"u}rgen Berges$^{1,3}$}
\author{Jens Pruschke$^{1}$}
\author{Alexander Rothkopf$^{2,3}$}
\affiliation{$^{1}$Institute for Nuclear Physics, Darmstadt University of Technology, Schlossgartenstr.\ 9, 64289 Darmstadt, Germany}
\affiliation{$^{2}$Department of Physics, University of Tokyo, Tokyo 113-0033, Japan}
\affiliation{$^{3}$Yukawa Institute of Theoretical Physics, University of Kyoto, Kyoto 606-8502, Japan}

\begin{abstract}
Nonequilibrium instabilities are known to lead to exponential
amplification of boson occupation numbers for low momentum modes on time
scales much shorter than the asymptotic thermal equilibration
time. We show for Yukawa-type interactions that this growth
induces very efficient fermion production, which proceeds with the maximum primary boson growth rate.
The description is based on a $1/N$
expansion of the 2PI effective action to NLO including boson-fermion 
loops, which are crucial to observe this phenomenon. 
For long enough amplification in the boson sector, fermion production terminates 
when the thermal occupancy is reached in the infrared. 
At higher momenta, where boson occupation numbers are
low, the fermion modes exhibit a power-law regime with
exponent two. 
\end{abstract}

\maketitle

\section{Introduction}
\label{sec:intro}

Fermion production in the presence of large boson occupation
numbers is an important question for various aspects of
early-universe cosmology as well as collision experiments
of heavy nuclei in the laboratory. Large boson occupation
numbers can lead to substantial deviations from perturbative
production processes. An important application concerns the
question of thermalization following nonequilibrium instabilities in the context of 
early-universe inflaton reheating 
dynamics~\cite{Traschen:1990sw,Kofman:1994rk,Boyanovsky:1994me,Khlebnikov:1996mc,Berges:2002cz}, or in quantum chromodynamics (QCD) relevant for relativistic heavy-ion collisions~\cite{Plasmainst,Plasmainst2,WongYangMills,Romatschke:2006nk,Berges:2007re,Fujii:2008dd}. Nonequilibrium instabilities
lead to exponential growth of boson occupation numbers in long
wavelength modes on time scales much shorter than the asymptotic
thermal equilibration time. This is followed by a period during
which bosons develop power cascades in a manner reminiscent of Kolmogorov wave 
turbulence~\cite{Micha:2002ey,Berges:2008wm,Barnaby:2009mc,Arnold:2005qs,Berges:2008mr}. For
long wavelengths a new class of nonperturbative scaling solutions 
has recently been found in the context of
early-universe reheating dynamics~\cite{Berges:2008wm}. The corresponding
nonthermal infrared fixed points~\cite{Berges:2008wm,Berges:2008sr} 
have the dramatic consequence
that a diverging time scale exists far from equilibrium, which can
prevent or substantially delay thermalization. Understanding these
phenomena in the presence of fermions is an important step towards
more realistic phenomenological applications.

In this work we study nonequilibrium fermion production in
Yukawa-type models following a spinodal/tachyonic instability.
More precisely, we consider a $SU(2)_L \times SU(2)_R$ symmetric
theory for massless Dirac fermions coupled to a $N=4$ vector of
scalars in $3+1$ dimensions. The model incorporates the chiral
symmetry of massless two-flavor QCD. It can also be considered as
a model for scalar inflaton dynamics coupled to fermions. For our purposes, a major
aspect of this model is that controlled
approximation schemes are available, which can describe the
far-from-equilibrium dynamics at early times as well as the
late-time approach to thermal equilibrium. It is based on a $1/N$
expansion to next-to-leading order (NLO) of the two-particle
irreducible (2PI) effective action out of equilibrium~\cite{Berges:2001fi,Berges:2002wr}. 
The fermion-boson loop-contributions included in this approximation
allow us for the first time to describe the phenomenon of instability-induced 
fermion production \footnote{The possibility of unstable fermionic modes was investigated in Refs.~\cite{Mrowczynski:2001az,Schenke:2006fz}
in an approximation, which does not include the mechanism discussed here.}: This production mechanism is based on the exponential growth of long-wavelength boson occupation numbers following a nonequilibrium instability. It is very efficient in the sense that the amplification of fermion occupation numbers proceeds with the maximum primary growth rate of boson occupancies. 
This can lead to a fast approach to
thermal equilibrium fermion distributions for low momentum modes,
though the bosons are still far from equilibrium at this stage. At
higher momenta, where boson occupation numbers are low, the
fermion occupancies exhibit a power-law regime with exponent two.
Apart from a numerical solution without further assumptions, we
present approximate analytical solutions which provide a detailed
understanding of the underlying dynamics.

Far-from-equilibrium fermion production is typically studied by integrating the Dirac 
equation in the presence of a classical macroscopic field, such as in 
Refs.~\cite{Boyanovsky:1995ema,Baacke:1998di,Greene:1998nh,Giudice:1999fb,GarciaBellido:2000dc,Peloso:2000hy,Tsujikawa:2000ik,BasteroGil:2000je,Nilles:2001ry,Baacke:2007ca} for scalar inflaton dynamics coupled to fermions, or as in 
Ref.~\cite{Gelis:2005pb}
for the production of quark pairs from classical gluon fields in the context of QCD.
Fermion dynamics in the presence  of inhomogeneous classical fields has been 
studied in an abelian Higgs model in 
1 + 1 dimensions~\cite{Aarts:1998td}. Analytical studies of perturbative dissipation and 
noise kernels for Yukawa-type models are presented in Ref.~\cite{Ramsey:1997fz}. Studies 
including quantum fluctuations, such that thermalization with the approach to 
Bose-Einstein and Fermi-Dirac distributions can be observed, have been performed in 
Refs.~\cite{Berges:2002wr,Berges:2004ce,Lindner:2007am} for the model considered in this 
work. Here we extend these studies by including the NLO corrections in the 2PI $1/N$ 
expansion and apply the approximation to the physics of nonequilibrium instabilities.
This has not been done before including fermion fluctuations. In contrast, the bosonic 
part of our model has been studied in great detail and the 2PI $1/N$ expansion to NLO is 
known to provide a quantitative description of nonperturbatively large boson fluctuations 
for the considered case with 
$N=4$~\cite{Berges:2002cz,Arrizabalaga:2004iw,Berges:2008wm}. Here we evaluate the 2PI 
effective action for vanishing macroscopic field. The observed phenomena in the fermionic 
sector are, therefore, entirely due to quantum corrections that are neglected in previous 
studies where the Dirac equation is solved in the presence of a classical macroscopic 
field. In particular, our findings of an induced exponential growth of fermion occupation 
number is distinct from the mechanism of resonant fermion production due to an 
oscillating inflaton field~\cite{Greene:1998nh}. Similarly, the observed power-law 
behavior at high momenta is shown to be a consequence of quantum fluctuations beyond 
mass-like corrections, which are absent for the considered massless fermions because of chiral symmetry. It will be an important further step beyond our current work to include 
non-vanishing macroscopic field values, in order to estimate the importance of quantum 
corrections to the Dirac equation in the presence of classical fields.  

The paper is organized as follows.
In the next section we introduce our model with its approximation and initial conditions.
In Sec.~\ref{sec:boson_dynamics} the boson dynamics is discussed, where we will describe the 
early-time behavior following the nonequilibrium instability as well as the subsequent 
evolution to a nonthermal fixed point.
Sec.~\ref{sec:fermion_dynamics} is dedicated to a detailed analysis of the fermion dynamics, with the
induced fermion production and the emergence of a power-law regime. 
In Sec.~\ref{sec:conclusions} we summarize and present our conclusions.
There are three Appendices. In the first we give numerical and analytical results for the 
time evolution of the spectral function.
Appendix~\ref{sec:LR_decomp} presents some details about the chiral decomposition of the fermion two-point 
correlation functions, and the last Appendix is about the employed numerical method.

\section{Model and approximation}
\label{sec:model_approx}

\subsection{$SU(2)_L \times SU(2)_R$ symmetric model}
\label{sec:model}

We consider a class of quantum field theoretical models for fermions interacting with scalar bosons. This type of models has been widely used in very different contexts, such as fermionic preheating after inflation or as effective theories for low-energy hadron physics in quantum chromodynamics (QCD). Here we consider a $SU(2)_L \times SU(2)_R \sim O(4)$ symmetric theory with $N_f = 2$ species of massless Dirac fermions coupled to $N_s = 4$ scalars. The classical action reads
\begin{eqnarray}\label{class_action}
\nonumber
\!\!\! S&=&\int d^4x\left\lbrace\bar{\psi}i\gamma^{\mu}\partial_{\mu}\psi +\frac{g}{N_f}\bar{\psi}\left( \sigma +i\gamma_5\tau^l\pi^l\right)\psi \right. \\\nonumber
 &&\left.\hspace{0.7cm}+\frac{1}{2}\left[\partial_{\mu}\sigma \partial^{\mu}\sigma+\partial_{\mu}\pi^{l} \partial^{\mu}\pi^{l}\right]\right.\\
 &&\left.\hspace{0.7cm}-\frac{1}{2}m^2\left(\sigma^2+\pi^l\pi^l \right)-\frac{\lambda}{4!N_s}(\sigma^2+\pi^l\pi^l )^2   \right\rbrace ,
\end{eqnarray}
where $\tau^l$ $(l=1,2,3)$ are the standard Pauli matrices and
$\gamma^{\mu}$ $(\mu=0,1,2,3)$ denote the Dirac matrices. We employ a metric with components $\eta_{\mu\nu}= {\rm diag}(1,-1,-1,-1)$, where summation over repeated indices is implied. The fermion fields $\psi(x)$ with $x=\{x^0,{\bf x}\}$ and $\bar{\psi} = \psi^\dagger \gamma^0$ are coupled via the Yukawa interaction $g$ to the scalars $\phi = \left\{\sigma,\pi^1,\pi^2,\pi^3\right\}$, whose self-coupling is $\lambda$ and mass parameter $m$. 

Far-from-equilibrium dynamics as well as subsequent thermalization can be efficiently studied in this model using a resummed $1/N$ expansion beyond leading order  \cite{Berges:2001fi}, based on the two-particle irreducible ($2$PI) effective action \cite{Luttinger:1960ua,Cornwall:1974vz,Chou:1984es,Calzetta:1986cq,Berges:2004yj}. A major benefit of the $1/N$ approximation is the possibility to quantitatively address dynamics even in those cases where fluctuations become nonperturbative in the coupling, which will be relevant for the physics considered in this work. In nonequilibrium quantum field theory the $2$PI resummed $1/N$ expansion is required to cure the secularity problem of the standard ($1$PI) $1/N$ expansion, which allows one to study the late-time behavior of quantum fields~\cite{Berges:2001fi}. For the quantum theory with action (\ref{class_action}) the  relevant nonequilibrium evolution equations have been derived in Ref.~\cite{Berges:2002wr} to which we refer for further details. Here we summarize the derivation and give the equations which we use for our analysis.        

The most general 2PI effective action can be written as a functional of one-point and two-point correlation functions, such as $\langle \phi(x) \rangle$ and $\langle \phi(x) \phi(y) \rangle$ for scalars and similarly with fermion fields. Here the brackets $\langle \ldots \rangle$ denote traces over an initial density matrix, which is discussed below. We will always evaluate the functional for vanishing one-point functions, i.e.\
$\langle \phi \rangle = \langle \psi \rangle = \langle \bar{\psi} \rangle = 0$, and denote the nonvanishing two-point correlation functions by $G(x,y) = \langle \phi(x) \phi(y) \rangle$ and $D(x,y) = \langle \psi(x) \bar{\psi}(y) \rangle$. In this case the $2$PI effective action can be written as~\cite{Luttinger:1960ua,Cornwall:1974vz}
\begin{eqnarray}
&&\!\!\!\!\!\! \Gamma[G,D] \, = \, -i\text{Tr}\left(D_0^{-1}D\right)+\frac{i}{2}\text{Tr}\left(G_0^{-1}G\right)
\nonumber\\
&&\!\!\!\!\!\!
-i\text{Tr}\ln\left(D^{-1}\right)+\frac{i}{2}\text{Tr}\ln\left(G^{-1}\right)+\Gamma_{2}[G,D] + \text{const.} \quad \label{2pi_action}
 \end{eqnarray}
with the free scalar inverse propagator
\begin{equation}
i G^{-1}_{0,ab}(x,y) \,=\, - \left(\partial_\mu\partial^\mu + m^2 \right) \delta^{(4)}(x-y)\, \delta_{ab} 
\end{equation} 
for $a,b = 1,\ldots,N_s$ and the massless free inverse fermion propagator
\begin{equation}
i D^{-1}_{0,ij}(x,y) \,=\, i \gamma^\mu \partial_\mu \, \delta^{(4)}(x-y)\, \delta_{ij}
\end{equation} 
for $i,j = 1,\ldots,N_f$. The corresponding components of the full boson (fermion) propagator are denoted by $G_{ab}(x,y)$ ($D_{ij}(x,y)$). The traces in (\ref{2pi_action}) indicate integration over time along the closed time path~\cite{Schwinger:1960qe} and over three spatial coordinates as well as summation over internal indices. The functional $\Gamma_2[G,D]$ only contains contributions from two-particle irreducible diagrams with propagator lines associated to $G_{ab}(x,y)$ and $D_{ij}(x,y)$ \footnote{A diagram is said to be two-particle irreducible if it does not become disconnected by opening two lines.}.

The equations of motion for the full propagators are obtained from the stationarity conditions~\cite{Luttinger:1960ua,Cornwall:1974vz}
\begin{equation}\label{stat_cond}
\frac{\delta\Gamma[G,D]}{\delta D_{ij}(x,y)} \,=\, 0\, , \quad\frac{\delta\Gamma[G,D]}{\delta G_{ab}(x,y)}\, =\, 0 \, ,
\end{equation}
where we omit Dirac indices in the notation. Self-energies are defined by functional differentiation of the 2PI part $\Gamma_2[D,G]$ as
\begin{eqnarray}
\Sigma_{ij}(x,y) &=& -i\frac{\delta\Gamma_2[G,D]}{\delta D_{ji}(y,x)} \, ,
\label{self_en_fermion} \\
\Pi_{ab}(x,y) &=& 2i\frac{\delta\Gamma_2[G,D]}{\delta G_{ab}(x,y)}\, . \label{self_en_boson}
\end{eqnarray}
Because of $SU(2)_L \times SU(2)_R \sim O(4)$ symmetry one can always choose the propagators to be diagonal, such that $G_{ab} = G\, \delta_{ab}$, $D_{ij} = D\, \delta_{ij}$ and similarly for the self-energies. 

It is convenient to decompose the time-ordered two-point correlation functions $D(x,y)$ and $G(x,y)$ into a real and an imaginary part. For the fermion correlator we use~\cite{Berges:2002wr}
\begin{equation}\label{fermi_correl}
D(x,y) \,=\, F(x,y)-\frac{i}{2}\rho(x,y)\, \mbox{sign}(x^0-y^0) \, ,
\end{equation}
with the real statistical two-point function $F(x,y)$ and the spectral function $\rho(x,y)$. For fermions $F$ is determined by the expectation value of the commutator of two fermion field operators, while $\rho$ is described in terms of the anti-commutator of two fields~\cite{Berges:2004yj}. A similar decomposition can be done for the boson correlator with
\begin{equation}\label{bose_correl}
G(x,y) \,=\, F_\phi(x,y)-\frac{i}{2}\rho_\phi(x,y)\, \mbox{sign}(x^0-y^0) \, ,
\end{equation}
where, in contrast to the fermion case, $F_\phi$ is determined by the expectation value of the anti-commutator of two boson field operators and the spectral function $\rho_\phi$ by their commutator~\cite{Berges:2004yj}. Correspondingly, the decomposition identities for the self-energies (\ref{self_en_fermion}) and (\ref{self_en_boson}) read
\begin{eqnarray}\label{fermi_selfen_decomp}
 \Sigma(x,y) &=& C(x,y) - \frac{i}{2} A(x,y)\, \mbox{sign}(x^0-y^0) \, ,
 \\
 \Pi(x,y) &=& -i \Pi_{\rm local}(x)\, \delta^{(4)}(x-y) 
 \nonumber\\
 &+& \Pi_F(x,y) - \frac{i}{2} \Pi_\rho(x,y)\, \mbox{sign}(x^0-y^0) \, .
\end{eqnarray}
Here we take into account a possible space-time dependent local term 
$\Pi_{\rm local}(x)$ for the boson self-energy, which is not present for the fermion self-energy~\cite{Berges:2002wr}. 

We will consider spatially homogeneous and isotropic systems. The chiral symmetry of the model (\ref{class_action}) together with parity and
CP invariance imply that for the fermion two-point functions only the vector components 
\begin{eqnarray}
F^\mu_V \,=\, \frac{1}{4} {\rm tr} \left( \gamma^\mu F \right)
\, &,& \quad 
\rho^\mu_V \,=\, \frac{1}{4} {\rm tr} \left( \gamma^\mu \rho \right) \, ,
\\
C^\mu_V \,=\, \frac{1}{4} {\rm tr} \left( \gamma^\mu C \right)
\, &,& \quad 
A^\mu_V \,=\, \frac{1}{4} {\rm tr} \left( \gamma^\mu A \right) 
\end{eqnarray}  
are non-vanishing~\cite{Berges:2002wr}, where the trace acts in Dirac space. The vector components can be decomposed in spatial Fourier space as
\begin{equation} 
F^{\mu}_V(x,y) \, =\,  \int_{{\bf p}}  e^{i{\bf p}({\bf x}-{\bf y})} F^{\mu}_V(x^0,y^0;{\bf p}) ,
\end{equation}
where we use the abbreviation $\int_{{\bf p}} = \int d^3p/(2\pi)^3$, with 
\begin{eqnarray}
 F^{0}_V(x^0,y^0;{\bf p}) & = & F^{0}_V(x^0,y^0;|{\bf p}|) \, ,
\nonumber\\
{\bf F}_V(x^0,y^0;{\bf p}) & = & \frac{\bf p}{|{\bf p}|}\, F_V(x^0,y^0;|{\bf p}|)\, ,
\qquad \label{isotrop_con}
\end{eqnarray}
and equivalently for the other two-point functions. 

Denoting times as $x^0 = t$, $y^0 = t'$ etc.\ and setting the initial time to zero \footnote{Here we consider evolution equations for Gaussian initial conditions as described in Sec.~\ref{sec:initial}.}, the equations of motion (\ref{stat_cond}) for the fermion two-point functions $F_V^0$, $F_V$, $\rho_V^0$ and $\rho_V$ read~\cite{Berges:2002wr}:
\begin{eqnarray}
&& i \partial_t F_V^0(t,t';|{\bf p}|) \,=\, |{\bf p}|\,F_V(t,t';|{\bf p}|)
\nonumber\\
&&+\int\limits_{0}^{t}d t'' \left\{ A_V^0(t,t'';|{\bf p}|)\, F_V^0(t'',t';|{\bf p}|)\right.
\nonumber\\
&&\left.-A_V(t,t'';|{\bf p}|)\, F_V(t'',t';|{\bf p}|)\right\}
\nonumber\\
&&-\int\limits_{0}^{t'}d t'' \left\{ C_V^0(t,t'';|{\bf p}|)\, \rho_V^0(t'',t';|{\bf p}|)\right.
\nonumber\\
&&\left.-C_V(t,t'';|{\bf p}|)\, \rho_V(t'',t';|{\bf p}|)\right\},
\label{FV0_eom}\\[0.2cm]
&& i \partial_t F_V(t,t';|{\bf p}|) \,=\, |{\bf p}|\,F_V^0(t,t';|{\bf p}|)
\nonumber\\
&&+\int\limits_{0}^{t}d t'' \left\{ A_V^0(t,t'';|{\bf p}|)\, F_V(t'',t';|{\bf p}|)\right.
\nonumber\\
&&\left.-A_V(t,t'';|{\bf p}|)\, F_V^0(t'',t';|{\bf p}|)\right\}
\nonumber\\
&&-\int\limits_{0}^{t'}d t'' \left\{ C_V^0(t,t'';|{\bf p}|)\, \rho_V(t'',t';|{\bf p}|)\right.
\nonumber\\
&&\left.-C_V(t,t'';|{\bf p}|)\, \rho_V^0(t'',t';|{\bf p}|)\right\},
\label{FV_eom}\\[0.2cm]
&& i \partial_t \rho_V^0(t,t';|{\bf p}|) \,=\, |{\bf p}|\,\rho_V(t,t';|{\bf p}|)
\nonumber\\
&&+\int\limits_{t'}^{t}d t'' \left\{ A_V^0(t,t'';|{\bf p}|)\, \rho_V^0(t'',t';|{\bf p}|)\right.
\nonumber\\
&&\left.-A_V(t,t'';|{\bf p}|)\, \rho_V(t'',t';|{\bf p}|)\right\},
\label{rhoV0_eom}\\[0.2cm]
&& i \partial_t \rho_V(t,t';|{\bf p}|) \,=\, |{\bf p}|\,\rho_V^0(t,t';|{\bf p}|)
\nonumber\\
&&+\int\limits_{t'}^{t}d t'' \left\{ A_V^0(t,t'';|{\bf p}|)\, \rho_V(t'',t';|{\bf p}|)\right.
\nonumber\\
&&\left.-A_V(t,t'';|{\bf p}|)\, \rho_V^0(t'',t';|{\bf p}|)\right\}\,.
\label{rhoV_eom}
\end{eqnarray}
Similarly, the equations of motion for the boson two-point functions $F_{\phi}$ and $\rho_{\phi}$ are
\begin{eqnarray}
&& \left\{ \partial_t^2 +{\bf p}^2+M^2(t) \right\} F_{\phi}(t,t';|{\bf p}|)\, = \, 
\nonumber\\[0.2cm]
&& -\int\limits_{0}^{t}d t'' \,\Pi_\rho(t,t'';|{\bf p}|)\, F_{\phi}(t'',t';|{\bf p}|)
\nonumber\\
&& + \int\limits_{0}^{t'}d t'' \,\Pi_F(t,t'';|{\bf p}|)\, \rho_{\phi}(t'',t';|{\bf p}|)\, ,
\label{Fboson_eom}\\[0.3cm]
&& \left\{ \partial_t^2 +{\bf p}^2+M^2(t) \right\} \rho_{\phi}(t,t';|{\bf p}|) \,=\,
\nonumber\\[0.2cm]
&&-\int\limits_{t'}^{t}d t'' \,\Pi_\rho(t,t'';|{\bf p}|)\, \rho_{\phi}(t'',t';|{\bf p}|) \, ,
\label{rhoboson_eom}
\end{eqnarray}
with the mass-like term
\begin{equation}\label{eff_mass0}
M^2(t) \, =\, m^2 + \Pi_{\rm local}(t)\, ,
\end{equation}
which depends on time only because of the assumed spatial homogeneity. The above equations have to be regularized, which we will do by implementing a momentum cutoff following Ref.~\cite{Berges:2002wr} as described in Appendix~\ref{sec:numerical}.

One observes the following symmetry or anti-symmetry relations with respect to interchanging time arguments~\cite{Berges:2002wr}:  
\begin{eqnarray}
F_V(t,t';|{\bf p}|)&=& F_V(t',t;|{\bf p}|) \, ,
\nonumber\\
F_V^0(t,t';|{\bf p}|)&=&-F_V^0(t',t;|{\bf p}|) \, ,
\nonumber\\
\rho_V(t,t';|{\bf p}|)&=&-\rho_V(t',t;|{\bf p}|) \, ,
\nonumber\\
\rho_V^0(t,t';|{\bf p}|)&=&\rho_V^0(t',t;|{\bf p}|)\, .
\label{sym_rel_f}
\end{eqnarray}
For the Fourier coefficients of the boson two-point functions the corresponding relations are
\begin{eqnarray}
F_{\phi}(t,t';|{\bf p}|)&=&F_{\phi}(t',t;|{\bf p}|) \, ,
\nonumber\\
\rho_{\phi}(t,t';|{\bf p}|)&=&-\rho_{\phi}(t',t;|{\bf p}|)\, .
\label{sym_rel_b}
\end{eqnarray}

\subsection{Approximation}
\label{sec:approx}

\begin{figure}[t]
 \begin{center}
 \includegraphics[scale=0.25]{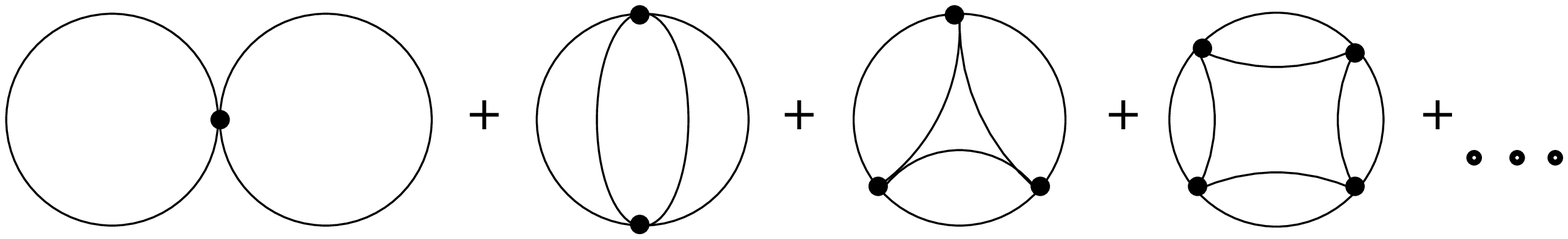}\\
\medskip
 \includegraphics[scale=0.35]{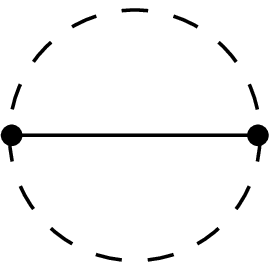}
\end{center}
\caption{Diagrammatic representation of the employed approximation for $\Gamma_2$. Solid lines are associated to self-consistently dressed boson propagators $G$ and dashed lines to fermion propagators $D$. The dots indicate an infinite series of diagrams, each obtained from the previous one by adding another loop in the bubble ring.}\label{fig:feyn_diag} 
\end{figure}

For given $\Gamma_2[G,D]$ the self-energies are expressed in terms of self-consistently dressed propagators $G$ and $D$ according to (\ref{self_en_fermion}) and (\ref{self_en_boson}). Here we consider the nonperturbative 2PI $1/N$ expansion to next-to-leading order (NLO) in the number of boson fields $N_s$~\cite{Berges:2001fi}. For bosons the physics of nonequilibrium instabilities cannot be described by an expansion in the coupling $\lambda$ even for weak couplings, since occupation numbers can grow  parametrically large of order $1/\lambda$~\cite{Berges:2002cz}. In contrast, occupation numbers for fermions are bounded due to the Pauli principle and a 2PI coupling expansion in powers of the Yukawa coupling $g$ will be employed to describe the fermion fluctuations. This requires to sum an infinite series of boson diagrams, which is displayed in Fig.~\ref{fig:feyn_diag} along with the single boson-fermion diagram contributing at this order.

The corresponding self-energies entering the evolution equations (\ref{FV0_eom})--(\ref{rhoboson_eom}) are for $N_s = 4$ and $N_f=2$~\cite{Berges:2002wr}
\begin{eqnarray}
\nonumber
C_V^{\mu}(t,t';{\bf p}) &=& -g^2 \int_{{\bf q}} \biggl\{F_V^{\mu}(t,t';{\bf q})F_{\phi}(t,t';{\bf p}-{\bf q})
\biggr.\\\label{CV_def}
&&\left.-\frac{1}{4}\rho_V^{\mu}(t,t';{\bf q})\rho_{\phi}(t,t';{\bf p}-{\bf q})\right\}\, ,\\
\nonumber
A_V^{\mu}(t,t';{\bf p}) &=& -g^2 \int_{{\bf q}} \biggl\{F_V^{\mu}(t,t';{\bf q})\rho_{\phi}(t,t';{\bf p}-{\bf q})
\biggr.\\\label{AV_def}
&&\biggl.+\rho_V^{\mu}(t,t';{\bf q})F_{\phi}(t,t';{\bf p}-{\bf q})\biggr\} 
\end{eqnarray}
for the fermion part and
\begin{eqnarray}
&&\Pi_{F}(t,t';{\bf p}) \, = \, -\frac{\lambda}{12} \int_{{\bf q}}
\biggl\lbrace F_{\phi}(t,t';{\bf p}-{\bf q})\biggr.
\nonumber\\
&& \biggl. \times\, I_{F}(t,t';{\bf q}) -\frac{1}{4}\rho_{\phi}(t,t';{\bf p}-{\bf q})I_{\rho}(t,t';{\bf q})\biggr\rbrace \nonumber\\
&&- 2 g^2 \int_{{\bf q}} \biggl\lbrace F^{\mu}_V(t,t';{\bf q})F_{V,\mu}(t,t';{\bf p}-{\bf q})\biggr.
\nonumber\\ &&\biggl.-\frac{1}{4}\rho^{\mu}_V(t,t';{\bf q})\rho_{V,\mu}(t,t';{\bf p}-{\bf q})\biggr\rbrace \, , 
\label{Sigma_F}\\
&& \Pi_{\rho}(t,t';{\bf p}) \, = \, -\frac{\lambda}{12} \int_{{\bf q}}
\biggl\lbrace F_{\phi}(t,t';{\bf p}-{\bf q})\biggr.
\nonumber\\ 
&&\biggl. \times\, I_{\rho}(t,t';{\bf q}) +\rho_{\phi}(t,t';{\bf p}-{\bf q})I_{F}(t,t';{\bf q})\biggr\rbrace 
\nonumber\\
&&- 4 g^2 \int_{{\bf q}} 
\biggl\lbrace \rho^{\mu}_V(t,t';{\bf q})F_{V,\mu}(t,t';{\bf p}-{\bf q})\biggr\rbrace  \label{Sigma_R}
\end{eqnarray}
for the boson part. Here the functions $I_{F}$ and $I_{\rho}$ contain the summation of the 
infinite series of boson loop diagrams at NLO:
\begin{eqnarray}
&& I_{F}(t,t';{\bf p}) \, = \, \frac{\lambda}{6}\int_{{\bf q}}\Bigg\lbrace F_\phi(t,t';{\bf p}-{\bf q}) F_\phi(t,t';{\bf q})
\nonumber\\
&& -\frac{1}{4}\rho_\phi(t,t';{\bf p}-{\bf q}) \rho_\phi(t,t';{\bf q})
- \int\limits_{0}^{t}dt''  I_{\rho}(t,t'';{\bf p})
\nonumber\\
&& \times \Bigg(F_\phi(t'',t';{\bf p}-{\bf q}) F_\phi(t'',t';{\bf q}) -\frac{1}{4} \rho_\phi(t'',t';{\bf p}-{\bf q})
\nonumber\\
&& \times\, \rho_\phi(t'',t';{\bf q})\Bigg) + 2\int\limits_{0}^{t'}dt''I_F(t,t'';{\bf p}) F_\phi(t'',t';{\bf p}-{\bf q})
\nonumber\\
&& \times \, \rho_\phi(t'',t';{\bf q})\Bigg\rbrace \, ,
\\
&& I_{\rho}(t,t';{\bf p}) \, = \, \frac{\lambda}{3} \int_{{\bf q}} \Bigg\{ F_\phi(t,t';{\bf p}-{\bf q})\rho_\phi(t,t';{\bf q})
\nonumber\\
&& - \int\limits_{t'}^{t} dt'' I_{\rho}(t,t'';{\bf p})F_\phi(t'',t';{\bf p}-{\bf q}) \rho_\phi(t'',t';{\bf q})
\Bigg\} \, .
\end{eqnarray} 
Finally, the mass-like term (\ref{eff_mass0}) reads
\begin{equation}\label{eff_mass}
M^2(t) = m^2 +  \frac{\lambda}{4} \int_{{\bf q}} F_{\phi}(t,t;{\bf q}).
\end{equation}
This approximation leads to a closed set of equations for the two-point functions $F_V^0$, $F_V$, $\rho_V^0$, $\rho_V$, $F_\phi$ and $\rho_\phi$.

\subsection{Initial conditions}
\label{sec:initial}

In order to solve the (integro-) differential equations of motion (\ref{FV0_eom})--(\ref{rhoboson_eom}) with the approximation (\ref{CV_def})--(\ref{eff_mass}), we have to specify initial conditions. The initial conditions for the spectral functions are fixed by their (anti-) commutation relations~\cite{Aarts:2001qa}, i.e.\ they are given by the equal-time properties for the fermion spectral function 
\begin{equation}\label{fermi_equal_t_prop}
 \rho^0_V(t,t;{\bf p}) \, = \, i \, ,\quad 
 \rho_V(t,t;{\bf p}) \, = \, 0  
\end{equation}
and for the boson spectral function 
\begin{equation}\label{bose_equal_t_prop}
\rho_{\phi}(t,t;{\bf p}) \, = \, 0\, ,\quad \partial_{t} \rho_{\phi}(t,t';{\bf p})\vert_{t=t'} \, = \, 1 \, , 
\end{equation}
which are valid for all times $t$. For the statistical functions we parametrize the Gaussian initial conditions as
\begin{eqnarray}
F^{0}_V(t,t';{\bf p})\vert_{t=t'=0} &=& 0  \, , \\
F_V(t,t';{\bf p})\vert_{t=t'=0} &=& \frac{1}{2} 
\label{FV_init}
\end{eqnarray}
for the fermions and 
\begin{eqnarray}\label{eq:bosoninitial_F}
 F_{\phi}(t,t';{\bf p})\vert_{t=t'=0} &=& \frac{1}{2 \epsilon_0({\bf p})} \, ,
 \label{init1_ferm}\\\label{eq:bosoninitial_dF}
\partial_{t}F_{\phi}(t,t';{\bf p})\vert_{t=t'=0} &=& 0 \, ,\\
\partial_{t}\partial_{t'}F_{\phi}(t,t';{\bf p})\vert_{t=t'=0} &=& \frac{\epsilon_0({\bf p})}{2}
\label{eq:bosoninitial_ddF}
\end{eqnarray}
for the bosons. Here $\epsilon_0({\bf p}) = \epsilon(t=0;{\bf p})$ plays the role of a quasi-particle energy at initial time, which we define using~\cite{Berges:2001fi} 
\begin{equation}\label{boson_disper} \epsilon(t,{\bf p})\equiv\sqrt{\frac{\partial_{t}\partial_{t'}F_{\phi}(t,t';{\bf p})}{F_{\phi}(t,t';{\bf p})}}\bigg\vert_{t=t'} \, .
\end{equation}
We note that the equal-time correlation function $F_V^0(t,t;|\textbf{p}|)$ corresponds to the conserved net charge density \cite{Berges:2002wr}, which vanishes identically at all times for our initial conditions.
We will consider the time-dependent quantities
\begin{eqnarray}
n_{\psi}(t,{\bf p}) & = & \frac{1}{2} - F_V(t,t;{\bf p}) \, ,
\label{eq:occferm}
\\
n_{\phi}(t,{\bf p}) & = & F_{\phi}(t,t;{\bf p})\, \epsilon(t,{\bf p}) - \frac{1}{2} \, ,
\label{eq:occbos}
\end{eqnarray}
which may be associated to effective fermion and boson occupation numbers, respectively.

In our simulations we start from a vacuum-like state with $n_{\psi}(t=0,{\bf p})= 0 $ and $n_{\phi}(t=0,{\bf p})=0$. We
induce a spinodal ("tachyonic") instability with a so-called quench by a sudden change in the sign in front of the $m^2$ term appearing in (\ref{eff_mass}). More precisely, we employ
\begin{eqnarray}
M^2(t=0) & = & m^2  , 
\nonumber\\
M^2(t>0) & = & - m^2
+ \frac{\lambda}{4} \int_{{\bf q}}  \left(F_\phi(t,t;{\bf q})-\frac{1}{2\epsilon_0({\bf q})}\right) , \qquad
\label{eq:ren_effmass}
\end{eqnarray}
where the constant subtraction of quadratically divergent terms in the integrand of (\ref{eq:ren_effmass}) ensures $M^2(t=0^+) = - m^2$ for the employed vacuum-like initial condition (\ref{init1_ferm}). We choose $m$ as our scale and express all quantities in appropriate powers of it.

Though all results that will be shown are obtained from the above initial conditions, we note that the evolution quickly becomes rather insensitive to the details at initial time. We will see that the spinodal instability leads to an exponential growth of fluctuations which soon dominate over the initial values of correlators. Late stages in the far-from-equilibrium evolution can even become universal in the sense that characteristic properties become insensitive to the details of the underlying microscopic field theory, which will be explained in the following.

\section{Boson dynamics}
\label{sec:boson_dynamics}

\subsection{Spinodal/tachyonic instability}
\label{sec:bose_instability}

\begin{figure}[t]
\includegraphics[scale=0.68]{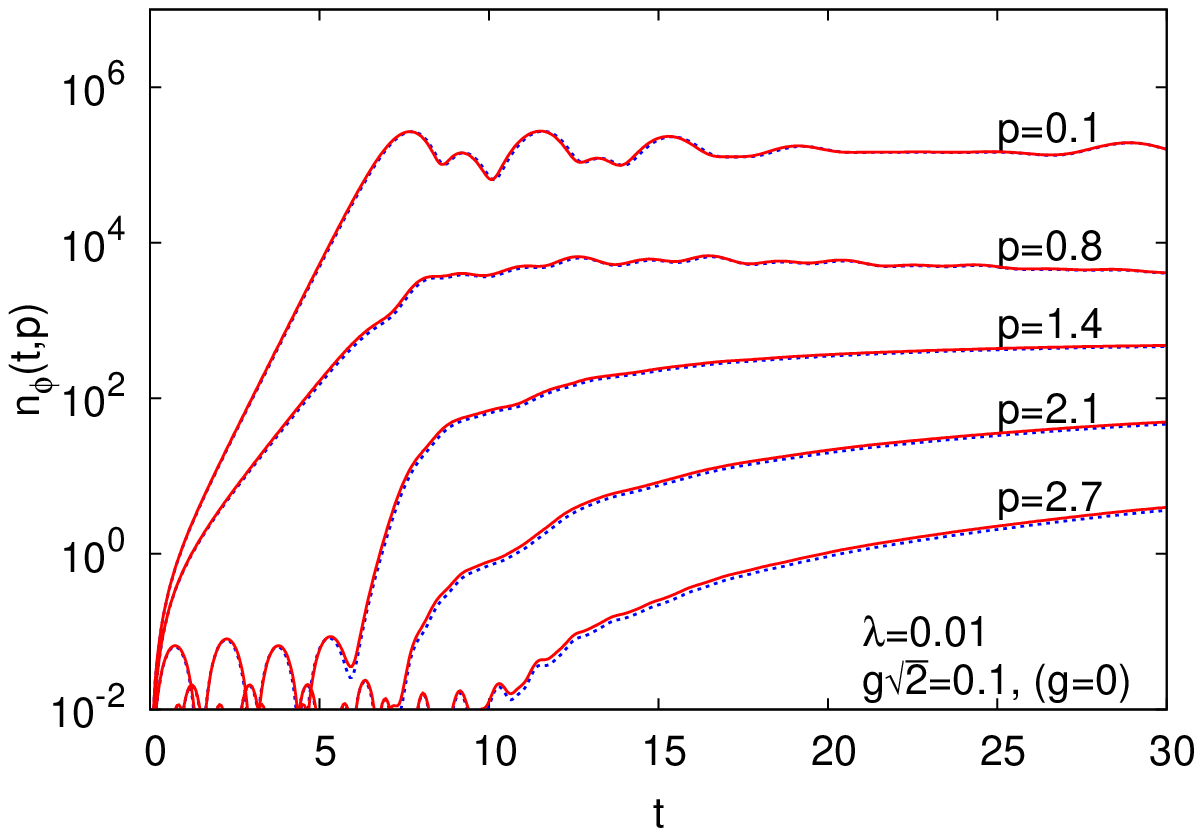}\\
\includegraphics[scale=0.68]{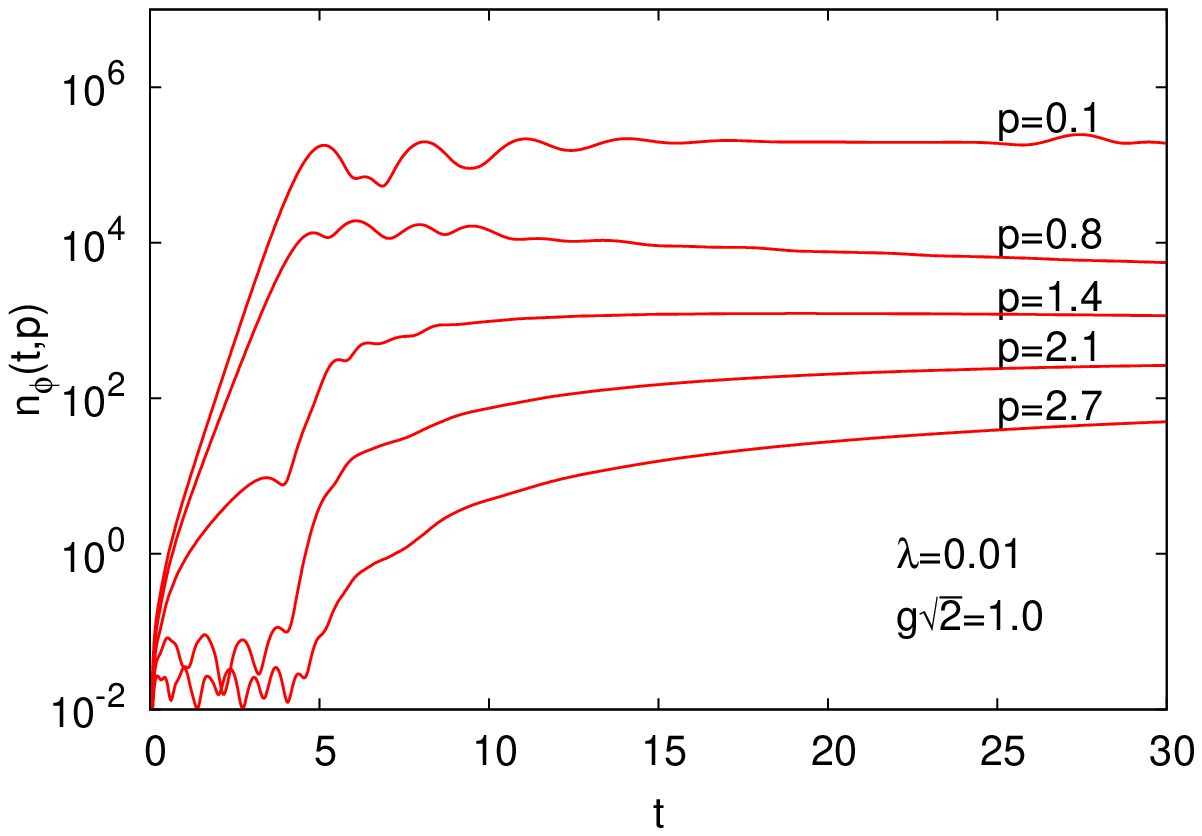}
\caption{Occupation numbers of boson modes as a function of time for given boson self-coupling $\lambda$ and different values of the Yukawa coupling $g$ to fermions. The upper graph shows the results for the purely bosonic theory with no fermions, $g=0$ (dashed lines), compared to a weakly coupled theory with $g\sqrt{2}=0.1$ (solid lines). The lower graph shows the behavior for $g\sqrt{2}=1$. Here and in all following figures the quantities are given in units of the mass parameter $m$.}\label{fig:nb_diffp_g01g0}
\end{figure} 
Since the evolution starts from a vacuum-like state, at very
early times almost no fluctuations are present for weak couplings. 
Accordingly, neglecting the self-energy corrections in (\ref{Fboson_eom}) and
(\ref{eff_mass0}) the initial evolution for the Fourier modes of the boson statistical propagator may be approximately described by 
\begin{equation}\label{eq:Fphi_linear_approx}
 \left[ \partial_t^2 + \textbf{p}^2 - m^2 \right] F_{\phi}(t,t';|{\bf p}|)\, \simeq \,0\, .
\end{equation}
Because of the negative sign in front of the $m^2 > 0$ term, the modes with $|\textbf{p}|<m$ exhibit exponential growth and the dominant solution behaves as
\begin{equation}
 F_\phi(t,t';\textbf{p}) \,=\, A_0\,  e^{\gamma(\textbf{p})\,(t+t')} \, ,
\label{Fphi_linear_approx_sol} 
\end{equation}
where the amplitude $A_0$ has the dimension of an inverse mass and we approximately have
$A_0 m \simeq 1/4$ for our initial conditions. The momentum dependent growth rate is given by
\begin{equation}
\gamma(\textbf{p}) = \sqrt{m^2-\textbf{p}^2} \, . 
\label{eq:growthrate}
\end{equation}
This classical instability leads to a strongest amplification with rate $\gamma_0 \equiv
\gamma(\textbf{p}=0)$ for the mode with the smallest momentum. This "primary" growth stage is characterized by a limited range of momenta $|\textbf{p}| < m$ for which amplification can be observed. During the subsequent exponential growth 
the fluctuations become larger and the self-energy corrections in (\ref{Fboson_eom}), (\ref{rhoboson_eom}) and (\ref{eff_mass0}) become relevant. These self-energy corrections incorporate non-linear effects, such as scattering and off-shell dynamics. For the bosonic model, in the absence of fermions, this has been discussed in detail for parametric resonance as well as spinodal instabilities in Refs.~\cite{Berges:2002cz,Berges:2004yj,Arrizabalaga:2004iw}. 

\begin{figure}[t]
\begin{center}
 \includegraphics[scale=0.68]{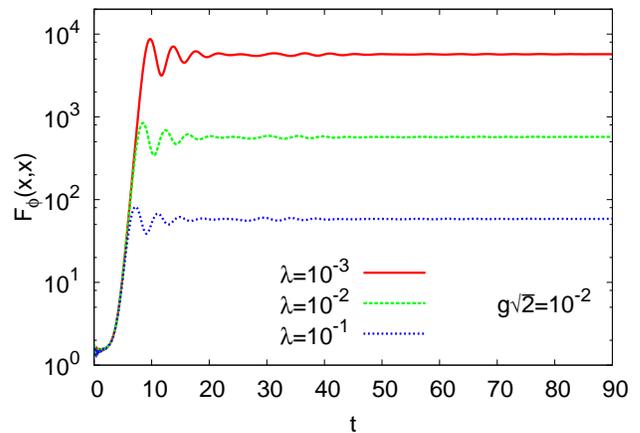}
\end{center}
\caption{Time evolution of the bosonic statistical two-point function $F_\phi(x,x)$ for different self-couplings and $g\sqrt{2}=10^{-2}$.}\label{fig:boseFxx}
\end{figure}
\begin{figure}[t]
\begin{center}
 \includegraphics[scale=0.68]{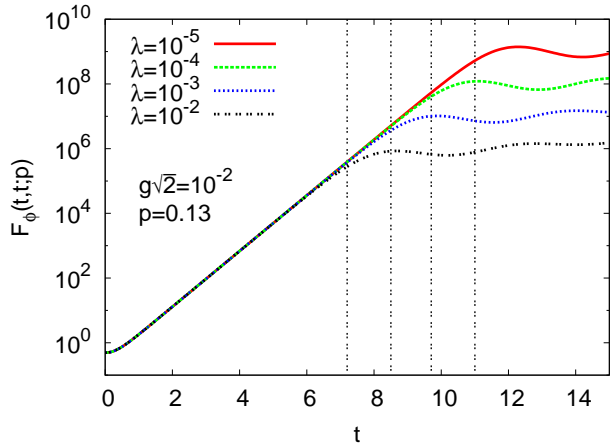}
\end{center}
\caption{Evolution of a low-momentum Fourier mode $F_\phi(t,t;|\textbf{p}|)$ for different couplings $\lambda$ and fixed $g$. The vertical dotted lines indicate the analytical estimates for the times at which the exponential growth period terminates: $t_\phi=7.2/m$, $8.5/m$, $9.7/m$, $11.0/m$ for $\lambda = 10^{-2}$, $10^{-3}$, $10^{-4}$, $10^{-5}$.}\label{fig:Fphi_tnonpert}
\end{figure}

In the upper graph of Fig.~\ref{fig:nb_diffp_g01g0} we show our results for the evolution of different boson occupation number modes (\ref{eq:occbos}) as a function of time for a scalar self-coupling $\lambda = 0.01$. The graph compares two runs, one without fermions ($g=0$) and the other with a weak Yukawa coupling of $g\sqrt{2}=0.1$. On the logarithmic plot one observes that the boson occupation numbers are practically not affected by the weak coupling to the fermions. The time evolution exhibits the characteristic stages explained in detail in the purely bosonic studies of Refs.~\cite{Berges:2002cz,Berges:2004yj,Arrizabalaga:2004iw}. 
Low-momentum modes with $|\textbf{p}| < m$ show primary exponential growth, whereas higher momentum modes become exponentially amplified at a later ("secondary") stage with faster growth rates due to nonlinearities that were built up by primary modes. After the fast period of exponential growth the dynamics slows down considerably. At this stage the nonlinearities are nonperturbatively large, i.e.~parametrically $F_\phi \sim {\cal O} (1/\lambda)$ and all processes become of order unity. This is illustrated for $F_\phi(x,x) = \int_{\bf p} F_\phi(t,t;{\bf p})$ for different values of the boson self-coupling in
Fig.~\ref{fig:boseFxx}. One observes that the ratio of this quantity taken at two coupling values, $\lambda$ and $\lambda'$, is well described by 
\begin{equation}
\frac{F_\phi(x,x)|_{\lambda}}{F_\phi(x,x)|_{\lambda'}} \, \simeq \, \frac{\lambda'}{\lambda}
\, .
\label{eq:Flambda}
\end{equation} 
The time at which this parametric dependence can be first observed we call $t_\phi$. This time was estimated in Ref.~\cite{Berges:2004yj} for the case of parametric resonance and we can adopt the method. More precisely, we consider the time when the local self-energy contribution (''tadpole'') of the mass-like term (\ref{eff_mass}) becomes one:
\begin{equation}
\frac{\lambda}{4} \int_{\bf q} F_\phi(t,t;{\bf q}) \,\simeq\, \frac{\lambda A_0\,e^{2 \gamma_0 t}}{32(\gamma_0^{-1}\,\pi\,t)^{3/2}} \, \stackrel{t=t_\phi}{=} 1\, .\label{tadpole_int}
\end{equation}
For the evaluation of the momentum integral we took (\ref{Fphi_linear_approx_sol}) and applied a saddle-point approximation around $|\textbf{p}|=0$. The last equality of (\ref{tadpole_int}) yields
\begin{equation}\label{eq:t_phi}
 t_\phi \, \simeq \, \frac{1}{2\gamma_0}\ln\left(\frac{32}{\lambda} \right)
+ \frac{3}{4\gamma_0}\ln\left(\frac{\pi\,t_\phi}{A_0^{2/3}\,\gamma_0^{-1/3}}\right) \, .
\end{equation}
Plugging in $A_0 m \simeq 1/4$ and using $\gamma_0=m$ one finds from the solutions of (\ref{eq:t_phi}) that $t_\phi\simeq 7.2/m$ for $\lambda=10^{-2}$, $t_\phi\simeq 8.5/m$ for $\lambda=10^{-3}$, $t_\phi\simeq 9.7/m$ for $\lambda=10^{-4}$ and $t_\phi\simeq 11.0/m$ for $\lambda=10^{-5}$. Fig.~\ref{fig:Fphi_tnonpert} shows a comparison of the analytical estimates with the numerical solutions for the nonequilibrium time evolution.
 
\begin{figure*}
 \begin{center}
\begin{large}\underline{$\lambda < g\,\,\,(2g^2/\lambda=1)$}       \end{large}\\
\includegraphics[scale=0.7]{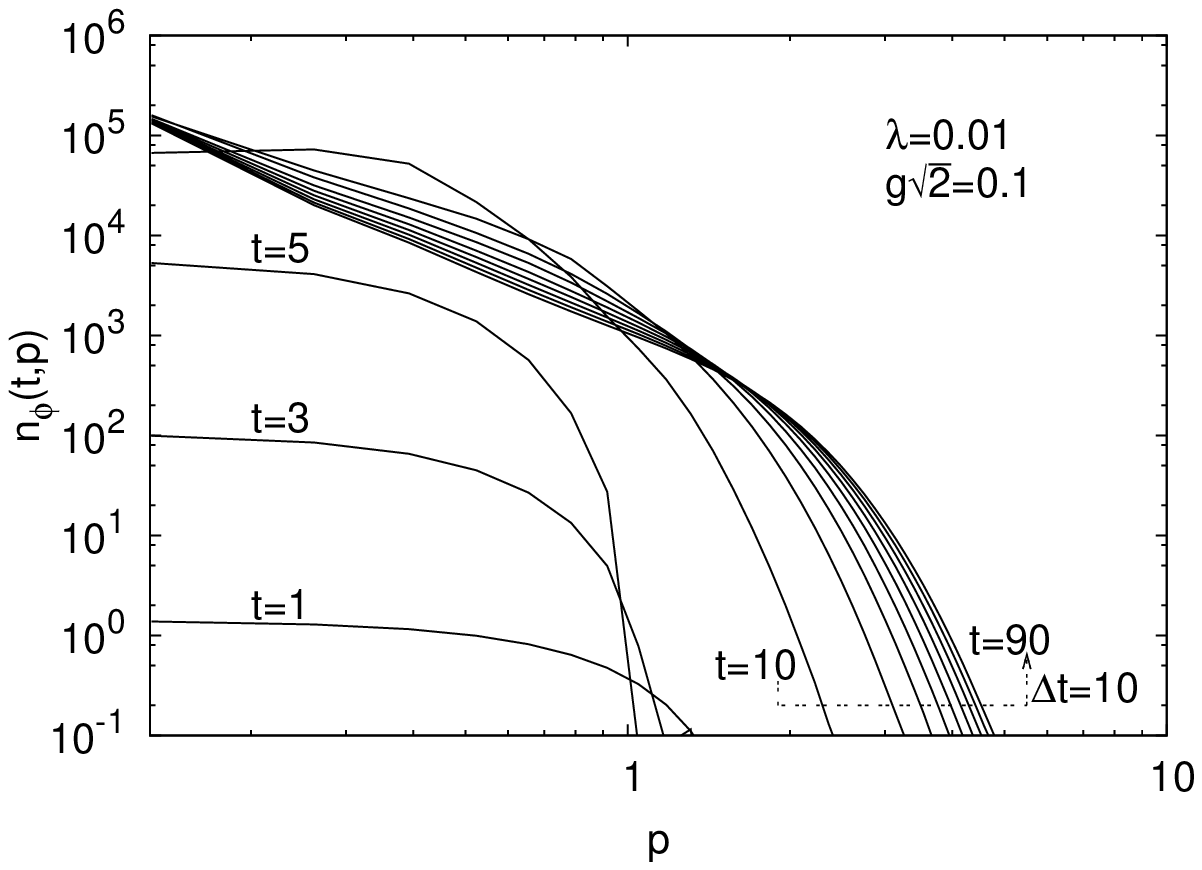}
\includegraphics[scale=0.7]{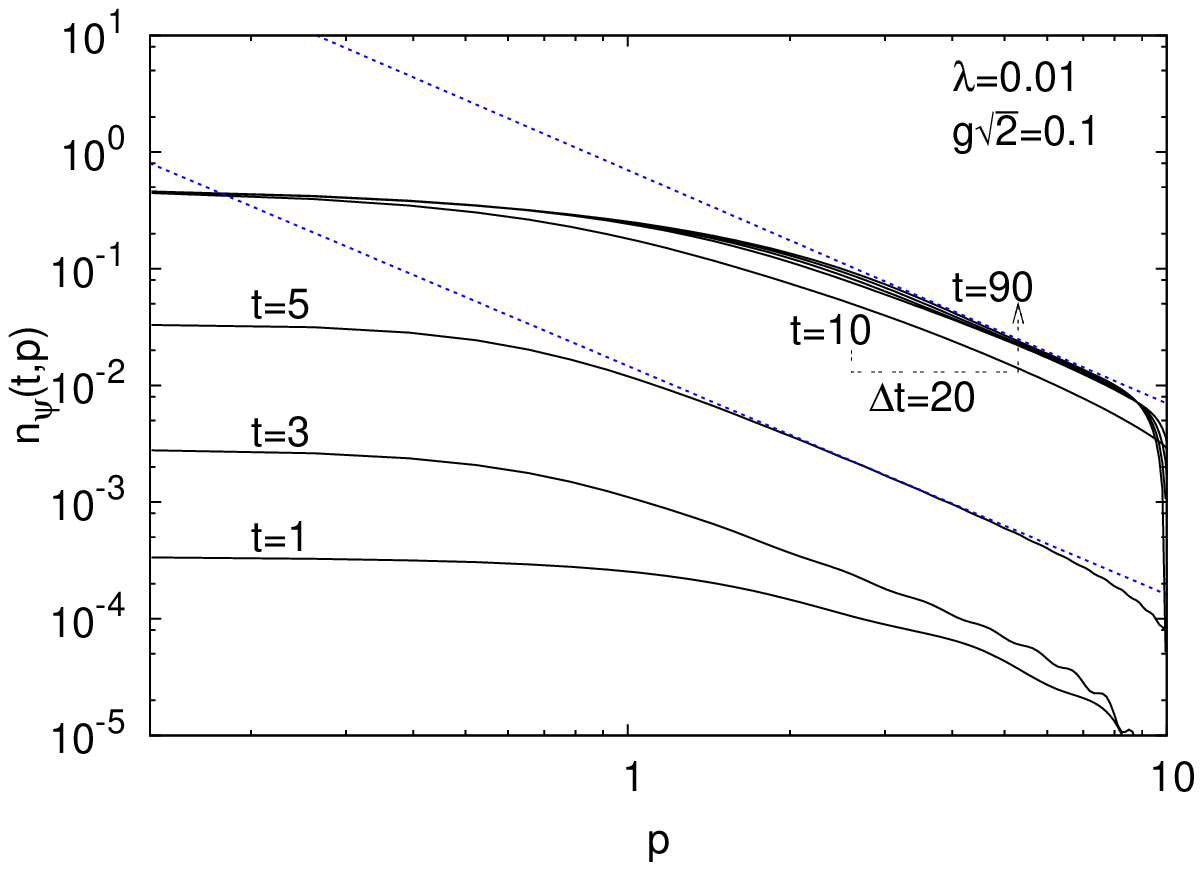}\\
\begin{large}\underline{$\lambda = g\,\,\,(2g^2/\lambda=10^{-2})$} \end{large}\\
\includegraphics[scale=0.7]{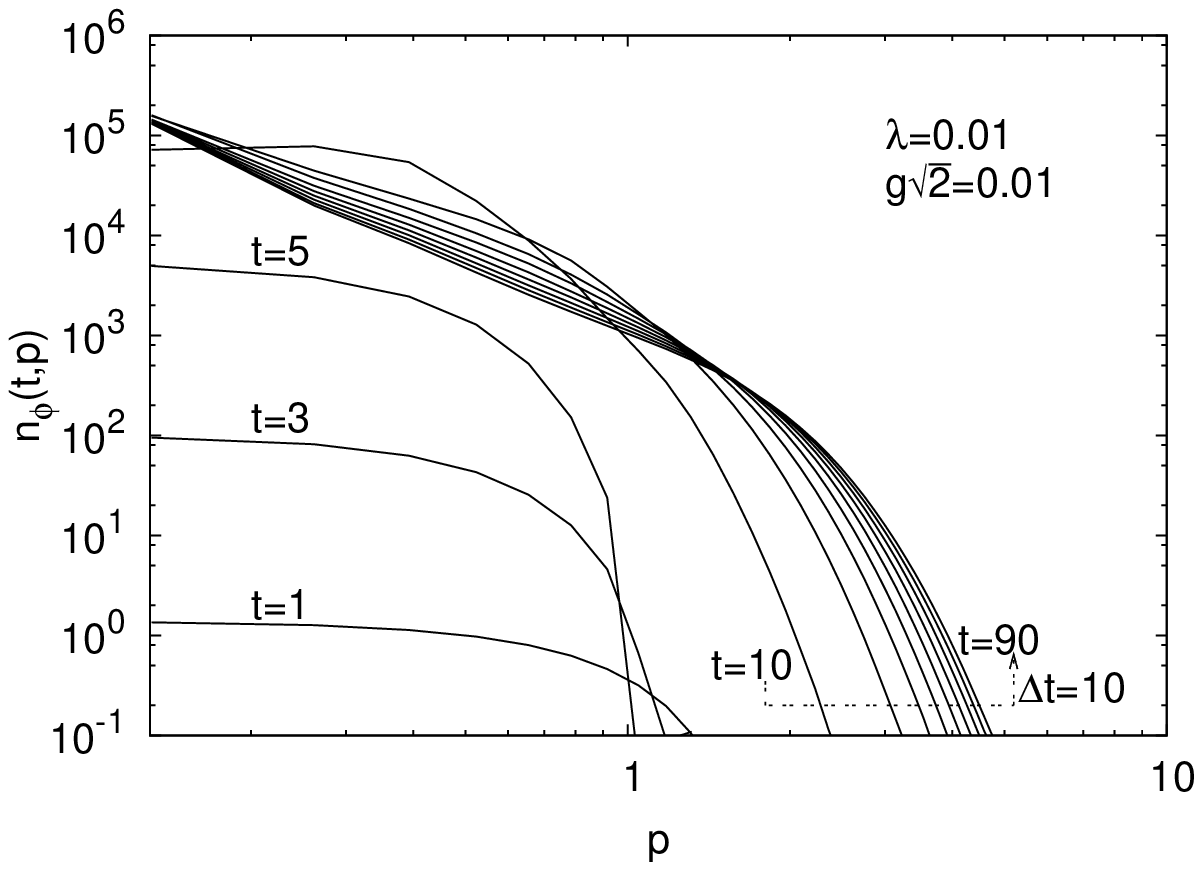}
\includegraphics[scale=0.7]{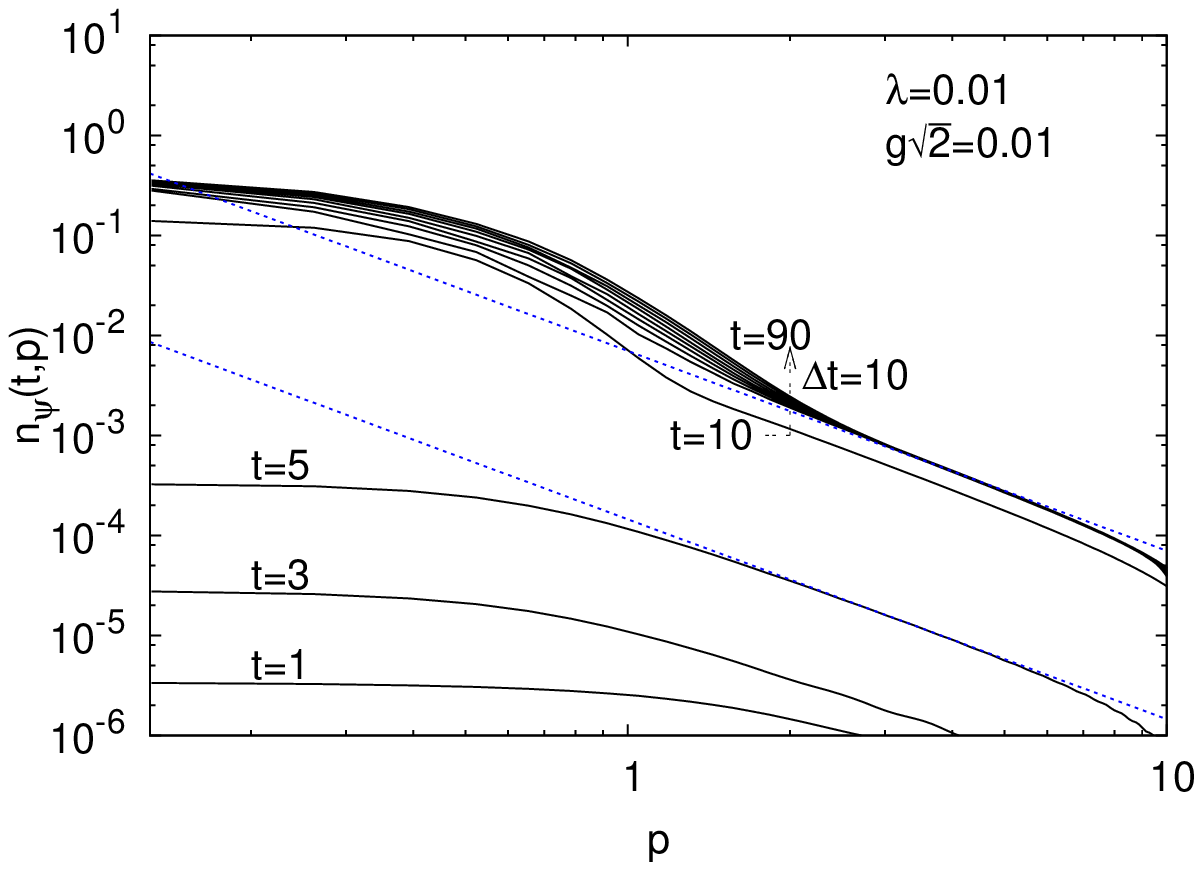}\\
\begin{large}\underline{$\lambda > g\,\,\,(2g^2/\lambda=10^{-3})$} \end{large}\\
\includegraphics[scale=0.7]{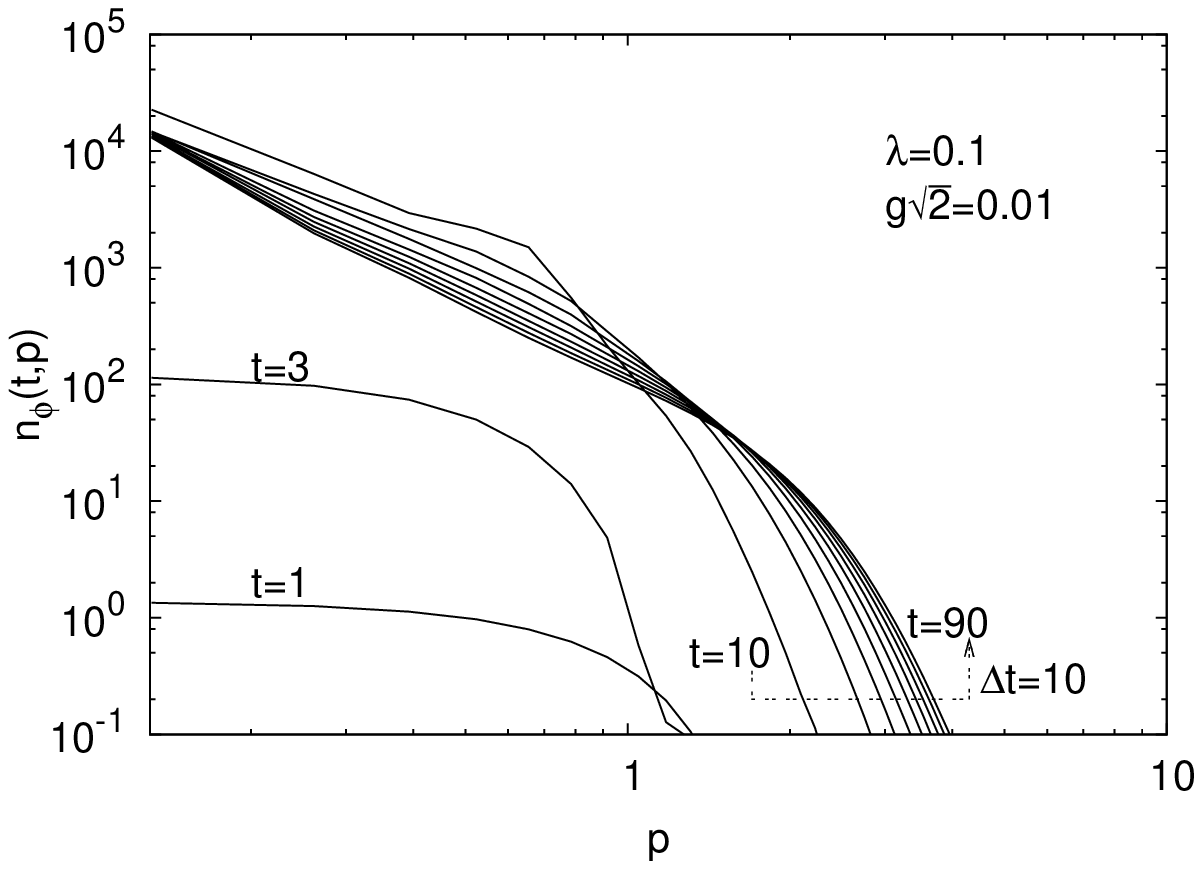}
\includegraphics[scale=0.7]{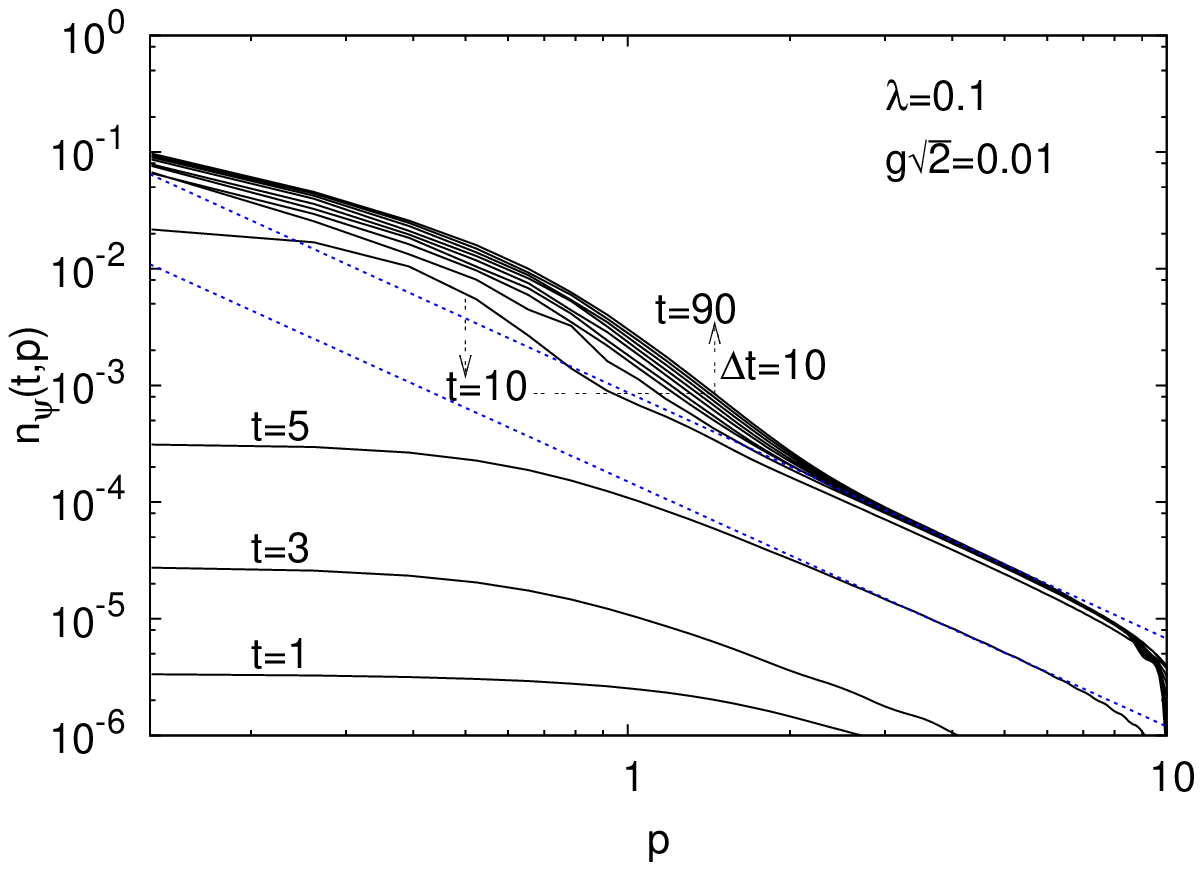}
\end{center}
\caption{Occupation numbers for bosons (left) and fermions (right) as a function of momentum at different times. The graphs from top to bottom correspond to different couplings as indicated in the figures. The dotted lines in the graphs for the fermion distributions display power-law behavior $n_\psi(|{\bf p}|) \sim |{\bf p}|^{-\kappa_{\psi}}$ with the exponent $\kappa_{\psi}=2$ for comparison (see Sec.~\ref{sec:fermion_dynamics}).}\label{fig:spectra_comp}
\end{figure*}

Occupation numbers as a function of momentum for different times are shown in Fig.~\ref{fig:spectra_comp}. The left column of that figure displays boson distributions for different couplings as indicated in the figure. The right column shows the corresponding behavior for the fermions, which will be discussed below in Sec.~\ref{sec:fermion_dynamics}. The time evolution of the boson occupation numbers is characterized by the exponential primary growth of low-momentum modes, and a subsequent broadening of the distributions as a consequence of enhanced secondary growth rates.   
Even though the Yukawa coupling changes one order of magnitude from $g\sqrt{2}=0.1$ to $g\sqrt{2}=0.01$, the corresponding graphs with scalar self-coupling $\lambda=0.01$ do not show
any significant differences. This confirms that weakly coupled fermions have no significant affect on the boson dynamics.

\begin{figure}[t]
\includegraphics[scale=0.34,angle=-90]{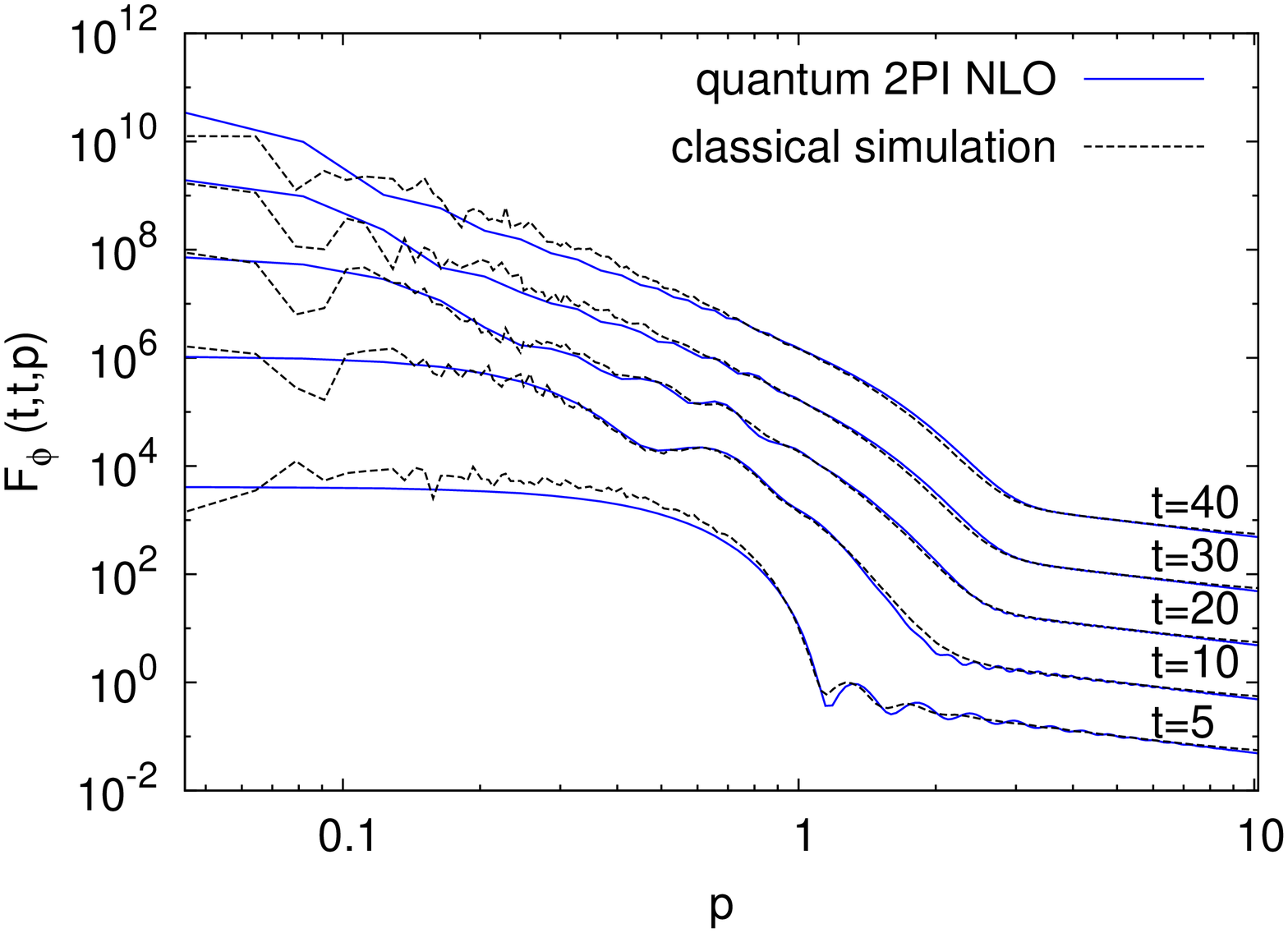}
\caption{Comparison between spectra of the boson correlation function $F_{\phi}(t,t;|{\bf p}|)$ in a 2PI (solid 
lines) and a classical-statistical (dotted lines) simulation at different times for $\lambda = 0.1$. The spectra for times larger than $t=5/m$ 
are multiplied by a factor of 10 in order to make separate lines better visible.}\label{fig:2PI_vs_class}
\end{figure}

If only the boson dynamics is considered without fermions, alternative nonperturbative approximation schemes are available that can be used to test the 2PI $1/N$ expansion~\cite{Aarts:2001yn,Arrizabalaga:2004iw,Berges:2007ym,Berges:2008wm}. 
Since the exponential growth is induced by an instability of the free-field-type equation (\ref{eq:Fphi_linear_approx}) at sufficiently early times quantum corrections may be neglected.  Furthermore, occupation numbers become large at later times because of the exponential growth and one expects an accurate description of the quantum dynamics using the classical-statistical field theory approximation for highly occupied modes. The classical-statistical simulation result is obtained from repeatedly solving the classical field equation of motion numerically and Monte Carlo sampling of the initial conditions. Fig.~\ref{fig:2PI_vs_class} shows a comparison of the purely boson dynamics, represented
by $F_\phi(t,t;|{\bf p}|)$, from the quantum 2PI $1/N$ expansion and the classical-statistical simulation for the initial conditions (\ref{eq:bosoninitial_F})-(\ref{eq:bosoninitial_ddF}) and $\lambda = 0.1$. 
The level of agreement between both calculations is remarkable. For low momenta, where occupation numbers are high, the classical-statistical simulation is expected to be accurate up to controlled statistical errors and the comparison points out the ability of the NLO approximation to describe the strongly nonlinear dynamics. On the other hand, from the comparison one also observes that for the considered small coupling the quantum corrections are small even for lower-occupied modes at higher momenta for not too late times.
For the asymptotic approach to quantum thermal equilibrium characterized by a Bose-Einstein distribution the classical-statistical approximation becomes invalid, while the quantum evolution using the 2PI $1/N$ expansion to NLO approaches a Bose-Einstein distribution~\cite{Berges:2001fi,Aarts:2001yn,Arrizabalaga:2004iw,Berges:2007ym}.

\subsection{Subsequent power-law behavior}
\label{sec:subsequent_pl}

\begin{figure}[t]
 \includegraphics[angle=-90,scale=0.34]{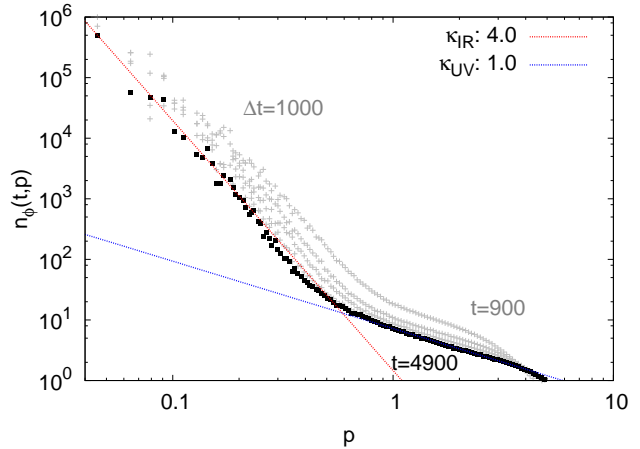}
\caption{The boson occupation numbers $n_\phi(t,|\textbf{p}|)$ from classical-statistical simulations for $\lambda=0.1$. Shown are in gray the distributions starting at time $t = 900/m$ (from right), continuing with a time step of $\Delta t = 1000/m$ to the black dots which correspond to $t=4900/m$.}\label{fig:non-them-fixed}
\end{figure}

The approach to thermal equilibrium can be substantially delayed in the presence of a nonequilibrium instability~\cite{Micha:2002ey}. This happens because after the exponential growth period the evolution is driven towards a nonthermal infrared fixed point. This was pointed out in Ref.~\cite{Berges:2008wm} for the case of a parametric resonance instability and discussed in terms of nonthermal renormalization group fixed points in Ref.~\cite{Berges:2008sr}. In order to demonstrate that a similar behavior also occurs starting from a spinodal or tachyonic instability, we present in Fig.~\ref{fig:non-them-fixed} classical-statistical simulation results for much later times than in Fig.~\ref{fig:2PI_vs_class}. Shown are snapshots of the occupation number as a function of momenta for different times, where the black dots are taken at the latest time $t = 4900/m$. A straight line on the double-logarithmic plot corresponds to a power-law behavior, 
\begin{equation}
n_{\phi}(t,|\textbf{p}|) \sim \frac{1}{|\textbf{p}|^{\kappa}} \, ,
\end{equation}
where the occupation number exponent $\kappa$ at low-momenta is well approximated by
$\kappa_{\rm IR} \simeq 4$ in agreement with the analytical estimates in Refs.~\cite{Berges:2008wm,Berges:2008sr} for the nonthermal fixed point as well as the numerical results starting from parametric resonance~\cite{Berges:2008wm}. At higher momenta, where parametrically $n_\phi(t,|\textbf{p}|) \lesssim 1/\lambda$, one observes the transition to a classical-statistical thermal distribution for which the occupation number exponent becomes $\kappa_{\rm UV} \simeq 1$. Towards higher momenta than those shown in Fig.~\ref{fig:non-them-fixed} the distribution still drops rapidly. The classical-statistical theory will finally occupy all high-momentum modes according to a power-law with $\kappa_{\rm UV} \simeq 1$, which corresponds to the Rayleigh-Jeans ultraviolet divergence. In contrast, quantum corrections in the 2PI $1/N$ approximation prevent a power-law distribution at high momenta and lead to finite results~\cite{Berges:2001fi,Aarts:2001yn,Arrizabalaga:2004iw,Berges:2007ym,Berges:2008sr}.
 
If the Yukawa coupling $g$ is increased in the presence of fermions, the dynamics of the bosons becomes substantially affected and their behavior can no longer be understood separately. This is demonstrated in the lower graph of Fig.~\ref{fig:nb_diffp_g01g0} for $g\sqrt{2}=1$. We emphasize that for $g \approx 1$ it may not be quantitatively justified to neglect higher loop-orders including fermion propagators beyond the two-loop graph taken into account in $\Gamma_2$ as shown in Fig.~\ref{fig:feyn_diag}. However, the observed increase of the growth rates comes from a negative mass-squared contribution for the boson self-energies, which is induced by fermion fluctuations. The very same mechanism operates already for small couplings with $g \ll 1$ and just the quantitative corrections are not sizeable enough in this case to be clearly visible in the figures. The induced negative mass-squared contribution is a well-known phenomenon, which has been analyzed in great detail also in the context of nonperturbative renormalization group studies of this model in thermal equilibrium~\cite{Berges:1997eu}. Even though the boson dynamics is not much influenced by a weak coupling to fermions and sizeable effects require stronger couplings, a similar observation can not be made for the fermion dynamics. It turns out that the latter is always substantially affected by the highly occupied boson modes, since even for weak Yukawa couplings they induce an exponential growth in the fermion correlation functions as is explained in the following.  

\section{Fermion dynamics}
\label{sec:fermion_dynamics}

\subsection{Instability-induced fermion production}
\label{sec:fermi_instability}

\begin{figure}[t]
\includegraphics[scale=0.68]{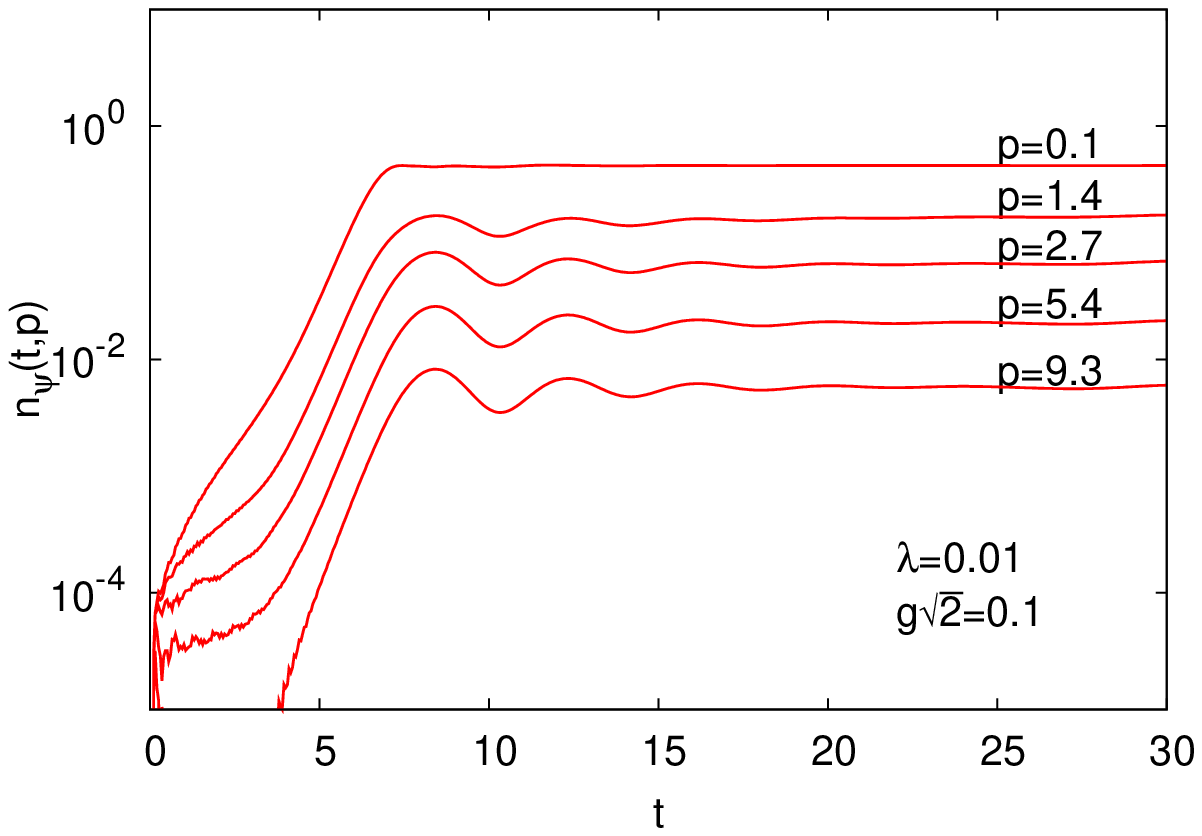}
\includegraphics[scale=0.68]{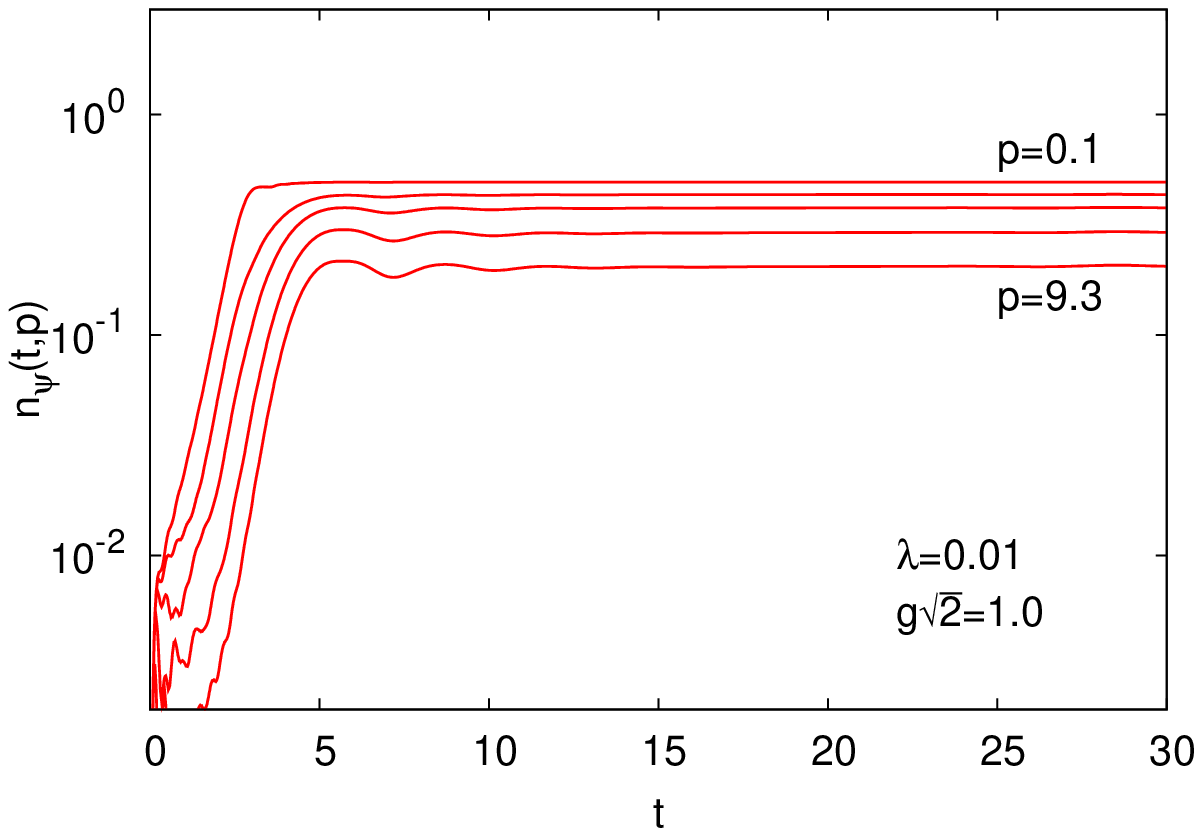}
\caption{Occupation numbers of fermion modes as a function of time with couplings as indicated in the graphs. The observed growth rate of modes is approximately independent of momenta. In the lower graph only two modes are labeled, the three other modes correspond to the same momenta as in the upper graph.}\label{fig:nf_diffp}
\end{figure} 
In Fig.~\ref{fig:nf_diffp} different modes of the fermion occupation number (\ref{eq:occferm}) are presented as a function of time for two different Yukawa couplings, $g\sqrt{2}=0.1$ (upper graph) and $g\sqrt{2}=1$ (lower graph), for $\lambda = 0.01$. The figure shows the behavior of the fermions, which corresponds to the evolution of the bosons presented in Fig.~\ref{fig:nb_diffp_g01g0}. After a short initial period 
one observes an exponential growth of fermion modes. The fast dynamics then slows down rather abruptly and approaches a quasi-stationary regime. 
A most characteristic property of the fast period is that the different momentum modes get exponentially amplified with approximately the same growth rate, which can be seen from the slopes of the curves in Fig.~\ref{fig:nf_diffp} for a wide momentum range $|{\bf p}|/m = 0.1$ -- $9.3$. This is in contrast to the amplification of boson modes, where the primary growth rates show a significant momentum dependence according to (\ref{eq:growthrate}) and the secondary rates equal multiples of the maximum primary growth rate (see Fig.~\ref{fig:nb_diffp_g01g0}). Similar to the behavior of the boson secondaries, one observes that higher-momentum fermion modes are amplified at later times. Though fermion production occurs practically over the whole spectrum, this delay leads to a decrease of the amplitude of the occupation number modes with increasing momenta. One also notes that in our simulations $n_\psi(t,\textbf{p})$ never exceeds $1/2$, despite the fact that its definition allows occupancies in the range $0\leq n_\psi(t,\textbf{p}) \leq 1$ in accordance with the Pauli principle. We note that $n_{\rm FD}(t,\textbf{p}=0) = 1/2$ corresponds to the value of the low-momentum distribution in thermal equilibrium, which is discussed in more detail in Sec.~\ref{sec:fermi_pl}. In the lower graph of Fig.~\ref{fig:nf_diffp} we show the same evolution but with a stronger Yukawa coupling $g\sqrt{2}=1$. One observes a more efficient fermion production over the spectrum, i.e.\ the decrease of the amplitude with higher momentum is less pronounced than for weak couplings. The growth rate is larger and the system enters the quasi-stationary evolution regime earlier. Corresponding plots for occupation numbers as a function of momentum for different times are shown in Fig.~\ref{fig:spectra_comp} above. The right column of that figure displays the fermion distributions for different couplings as indicated in the figure.
The emergence of the characteristic power-law behavior for higher momenta seen in Fig.~\ref{fig:spectra_comp} will be discussed below in Sec.~\ref{sec:fermi_pl}.  

\begin{figure}[t]
 \begin{center}
 \includegraphics[scale=0.5]{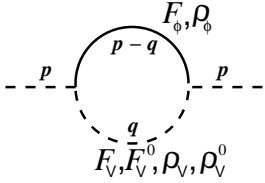}
\end{center}
\caption{Topology of the fermion self-energy diagram. The internal lines represent the various propagators indicated in the diagram.}\label{fig:fermi_loop_eom}
\end{figure}

In the following we derive an approximate analytical solution for the fermion dynamics, which explains the mechanism of the observed exponential growth and major characteristics. For weak Yukawa coupling the evolution of the boson correlation functions can be established without taking into account the fermion dynamics as discussed above. This is our starting point for an approximate evaluation of the fermion dynamics. We consider the equation of motion (\ref{FV_eom}) and concentrate on the equal-time correlator, $F_V(t,t;\textbf{p})$, since this determines the behavior of the effective particle number (\ref{eq:occferm}).
Because of the symmetry relations (\ref{sym_rel_f}) the term in (\ref{FV_eom}) proportional to the momentum $|\textbf{p}|$ vanishes. 
With the definition of the self-energies~(\ref{CV_def}) and~(\ref{AV_def}) there are eight terms, each coming with a product of two fermion and one boson two-point correlation function. The self-energy contributions can be visualized by a loop diagram as shown in Fig.~\ref{fig:fermi_loop_eom}, where internal lines represent the various propagators as indicated in the diagram. Because of the exponential growth of boson occupation numbers, the value of the respective equal-time correlation function $F_{\phi}(t,t;\textbf{p})$ becomes quickly many orders of magnitude larger than any other function. In particular, it becomes much larger than the spectral function $\rho_{\phi}(t,t';\textbf{p})$, which even vanishes at equal times. (See Appendix~\ref{sec:Fphi_dom} for a detailed discussion of the unequal-time properties of the spectral function.) This implies that all terms which include $F_{\phi}(t,t';\textbf{p})$ will dominate the self-energy corrections in (\ref{FV_eom}). The dominance is best realized if the boson self-coupling is small since the amplitude of $F_\phi$ becomes at the end of the exponential growth period parametrically of order $1/\lambda$ according to (\ref{eq:Flambda}). Hence our analytical estimates will require small $\lambda$. Neglecting the self-energy contributions without $F_{\phi}(t,t';\textbf{p})$ leads to four remaining terms and (\ref{FV_eom}) becomes for equal times
\begin{align}\nonumber
& i \left\{\partial_{t}\,F_V(t,t';|\textbf{p}|)\right\}\vert_{t'=t} \, \simeq \\\nonumber
&-g^2\!\!\int\limits_{0}^{t}\! dt''\!\!\int_{{\bf q}}\biggl[ \rho^0_V(t,t'';|{\bf q}|)F_{\phi}(t,t'';|{\bf p}-{\bf q}|)
F_V(t'',t;|\textbf{p}|)\biggr. \\\nonumber
&\biggl. -\frac{{\bf p}}{\vert{\bf p}\vert}\frac{{\bf q}}{\vert{\bf q}\vert}\rho_V(t,t'';|{\bf q}|)F_{\phi}(t,t'';|{\bf p}-{\bf q}|)
F^0_V(t'',t;|\textbf{p}|)\biggr]\\\nonumber
&+g^2\!\!\int\limits_{0}^{t}\! dt''\!\!\int_{{\bf q}} \biggl[ F^0_V(t,t'';|{\bf q}|)F_{\phi}(t,t'';|{\bf p}-{\bf q}|)
\rho_V(t'',t;|\textbf{p}|)\biggr. \\\label{eq:FV_eom_approx}
&\biggl.-\frac{{\bf p}}{\vert{\bf p}\vert}\frac{{\bf q}}{\vert{\bf q}\vert}F_V(t,t'';|{\bf q}|)F_{\phi}(t,t'';|{\bf p}-{\bf q}|)
\rho_V^0(t'',t;|\textbf{p}|)\biggr]\, .
\end{align}

Since we are considering the evolution of the equal-time correlation function, we employ 
for the derivative on the left hand side of (\ref{eq:FV_eom_approx}): 
\begin{eqnarray}\label{time_deriv}
\partial_t F_V(t,t;\textbf{p}) &=& \left\{\partial_{t'} F_V(t',t;\textbf{p})\right\}\vert_{t'=t} \nonumber\\
&& +\left\{\partial_{t'} F_V(t,t';\textbf{p})\right\}\vert_{t'=t} \nonumber\\
&=& 2 \left\{\partial_{t'} F_V(t',t;\textbf{p})\right\}\vert_{t'=t} \, ,
\end{eqnarray}
where we used in the second equation that $F_V(t',t;\textbf{p})$ is symmetric under time exchange according to (\ref{sym_rel_f}). 

For not too early times the integral in (\ref{eq:FV_eom_approx}) is dominated by momenta $\textbf{q} \simeq \textbf{p}$ for small enough $\textbf{p}$. To observe this we consider the primary growth rate (\ref{eq:growthrate}) for $\textbf{p} \ll m$, which is approximately given by
\begin{equation}
\gamma(\textbf{p}) \, \simeq\,  m  - \frac{\textbf{p}^2}{2 m} \, .
\label{eq:ratesmallp}
\end{equation}  
With (\ref{Fphi_linear_approx_sol}) and $\gamma_0 = m$ the low-momentum modes behave as
\begin{equation}
F_\phi(t,t;\textbf{p}-\textbf{q}) \,\simeq \, A_0\, e^{2 \gamma_0 t}\, e^{-\gamma_0^{-1} (\textbf{p}-\textbf{q})^2\, t} \, ,
\label{eq:Fphismallp} 
\end{equation}
such that the integrand of (\ref{eq:FV_eom_approx}) is dominated by $\textbf{q} \simeq \textbf{p}$ for sufficiently large $t$. This holds also in the presence of additional powers of momenta in the integrand, such as coming from the measure of the integral. The sum appearing in the integrand of (\ref{eq:FV_eom_approx}) consists of pairs of terms, which are the same if $\textbf{q} = \textbf{p}$. Accordingly, the equation of motion may be approximated by
\begin{align}\nonumber
 &i\partial_{t}F_V(t,t;|{\bf p}|) \, \simeq \,
- 4 g^2 \int\limits_{0}^{t}\! dt''\!\!\int_{{\bf q}}F_{\phi}(t,t'';|{\bf p}-{\bf q}|)\\
&\biggl[ \rho^0_V(t,t'';|{\bf p}|)F_V(t'',t;|\textbf{p}|)-\rho_V(t,t'';|{\bf p}|)F^0_V(t'',t;|\textbf{p}|)\biggr]. \label{eq:Fphiappeq}
\end{align}

During the time interval of exponential growth of $F_{\phi}(t,t';|\textbf{p}|)$ the latest times dominate the time- or "memory-integral" in (\ref{eq:Fphiappeq}). The characteristic time scale for these dominant contributions is given by the primary growth rate $\gamma_0$. Following similar calculations for boson dynamics in Refs.~\cite{Berges:2002cz,Berges:2004yj}, we will employ a memory expansion for an approximate analytical description of the fermion evolution. Integrals over the entire past, $\int_0^t dt'$, are replaced by integrals with a finite time range, $\int^t_{t- c/\gamma_0} dt'$. We will see that during the growth period this memory restriction leads to a good description of the full numerical results even for rather short memory intervals with $c\lesssim 1$. In particular, the estimate for the fermion growth rate will be insensitive to the value of $c$. Furthermore, we Taylor expand the integrand in (\ref{eq:Fphiappeq}) around the latest time of the memory integrals, where we keep only the lowest order in the expansion. Using the antisymmetry in time as described by (\ref{sym_rel_f}) and (\ref{sym_rel_b}) the equal-time correlation functions $F_V^0(t,t;|{\bf p}|)$ and  $\rho_V(t,t;|{\bf p}|)$ vanish,
such that the corresponding terms in (\ref{eq:Fphiappeq}) can be neglected to lowest order. Using (\ref{fermi_equal_t_prop}) we note that $\rho_V^0(t,t;|\textbf{p}|)=i$ for all momenta and times. Applying these approximations, the memory expanded evolution equation reads
\begin{equation}\label{FV_eq_final}
\partial_{t}F_V(t,t;|\textbf{p}|) + K(t) F_V(t,t;|\textbf{p}|) \, \simeq \, 0\, ,
\end{equation}
where we define
\begin{eqnarray}
 K(t) &=&  \frac{4 g^2 c}{\gamma_0} \int_{{\bf q}} F_{\phi}(t,t;|\textbf{p} - \textbf{q}|) 
 \nonumber\\
 &\simeq& g^2 \, \frac{A_0 \sqrt{\gamma_0} c}{2 (\pi t)^{3/2}}\, e^{2 \gamma_0 t} \,.
\label{def:K}
\end{eqnarray}
In order to perform the integral in (\ref{def:K}), we used (\ref{eq:Fphismallp}) and the fact that the contributions from high momenta are exponentially suppressed such that it may be well approximated by a Gaussian integral. We note that all of the above analytical approximations respect chiral symmetry, which is also discussed in more detail in Appendix~\ref{sec:LR_decomp}. We emphasize that the above approximations are not valid for $2 \gamma_0 t \ll 1$ since $F_\phi(t,t;|\textbf{p}|)$ is not large initially and one observes that $K(t)$ is not defined at $t=0$. However, we will see that even at early times the description is still reasonable and one can solve (\ref{FV_eq_final}) with the initial condition $F_V(t,t;|\textbf{p}|)\vert_{t=0^+}=1/2$ in accordance with (\ref{FV_init}).   

The solution of the equation of motion can be determined analytically and is given by
\begin{eqnarray}
&& F_V(t,t;|\textbf{p}|) \, = \,
\nonumber\\
&& \frac{1}{2} \exp\left\{ - g^2 a\, e^{2\,\gamma_0\,t} \left[2\, {\rm daw}(\sqrt{2 \gamma_0 t}) - \frac{1}{\sqrt{2 \gamma_0 t}} \right]\right\} \, , \qquad
\label{FVsol}
\end{eqnarray}
with the dimensionless constant
\begin{equation}
a \, = \, \frac{A_0 \gamma_0 \sqrt{2} c}{\pi^{3/2}} \, .
\label{eq:smalla}
\end{equation} 
The Dawson function appearing in (\ref{FVsol}) is defined as
\begin{equation}
{\rm daw}(x) \, = \, e^{-x^2} \int_0^x {\rm d} y\, e^{y^2} \,,
\end{equation}
which is closely related to the error function \cite{Abramowitz:1964}. It is an odd function and its derivative is
${\rm daw}'(x) \, =\, 1 - 2 x \, {\rm daw} (x)$
such that it grows linearly at $x=0$. Its maximum occurs at $x_0 \simeq 0.924$ and 
${\rm daw}(x_0) \simeq 0.541$. It has the asymptotic series
\begin{equation}
{\rm daw}(x) \, = \, \frac{1}{2 x} +  \frac{1}{2^2 x^3} + \ldots
\label{eq:dawsonasym}
\end{equation}
with which the expression (\ref{FVsol}) can be simplified for not too early times by using
\begin{equation}
2\, {\rm daw}(\sqrt{2 \gamma_0 t}) - \frac{1}{\sqrt{2 \gamma_0 t}} \, = \,
\frac{1}{2 (2 \gamma_0 t)^{3/2}} + \ldots\, .
\label{eq:asymptotic}
\end{equation}
One finds that (\ref{eq:asymptotic}) provides a rather good description already for times
$2\gamma_0 t \gtrsim \mathcal{O}(1)$.

\begin{figure}[t]
 \includegraphics[scale=0.68]{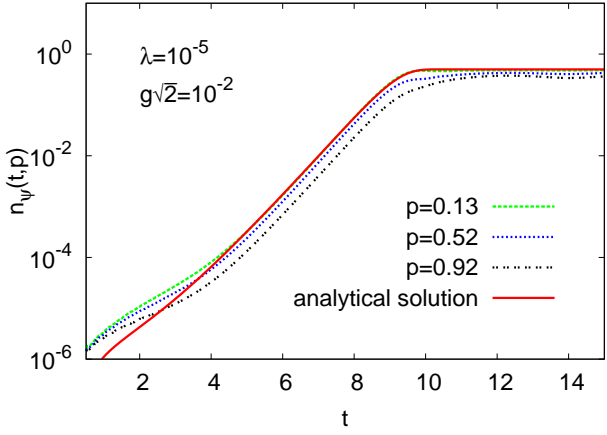}
\caption{Evolution of $n_{\psi}(t,|\textbf{p}|)$ as a function of time for different momenta and $g\sqrt{2}=10^{-2}$. The boson self-coupling is $\lambda=10^{-5}$. The solid line represents the analytic solution and the dotted lines show the results from simulations.}\label{fig:nf_analyt_diffp}
\end{figure}

The behavior of the fermion occupation number $n_{\phi}(t,|\textbf{p}|)$ is obtained by plugging the solution (\ref{FVsol}) into (\ref{eq:occferm}). With $A_0 \gamma_0 \simeq 1/4$ the constant (\ref{eq:smalla}) is approximately $a\simeq 0.05 c$. We typically find best agreement with the numerical solution for $c \simeq 0.48$, which is further discussed below.
A very characteristic feature of the solution (\ref{FVsol}) is that it predicts a growth rate which is independent of momenta. In Fig.~\ref{fig:nf_analyt_diffp} simulation results for $n_{\psi}(t,|\textbf{p}|)$ for different momenta are compared to the corresponding analytical solution for $g\sqrt{2}=10^{-2}$ and $\lambda=10^{-5}$. There are deviations at early times for the reasons mentioned above. In particular the asymptotic growth rate is accurately reproduced and indeed it is approximately momentum independent, which can also be seen from Fig.~\ref{fig:nf_diffp}. The quantitative agreement of the analytical and full numerical solution is good for small momenta as expected, with sizeable corrections already for $|\textbf{p}| \simeq m$. 

\begin{figure}[t]
\includegraphics[scale=0.68]{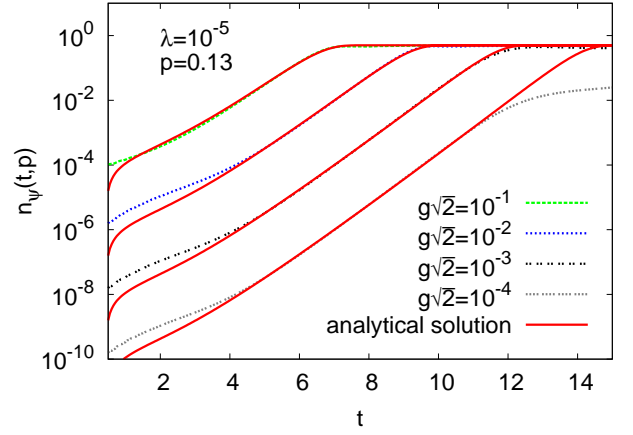}
\caption{Evolution of $n_{\psi}(t,|\textbf{p}|)$ as a function of time. The analytical solution (solid lines) is compared to the result from numerical simulations (dotted lines) for different values of the Yukawa coupling $g$. For the numerical simulations the lowest available momentum $|\textbf{p}|=0.13m$ is chosen.}\label{fig:nf_analyt_diffg}
\end{figure}

In the following we will discuss the solution (\ref{FVsol}) for two characteristic time ranges (I) $2 \gamma_0 t \gtrsim {\cal O}(1)$ and (II) $2 \gamma_0 t \gg 1$ for weak coupling $g \ll 1$. The r.h.s.\ of (\ref{FVsol}) contains two exponential functions. 
For the regime (I) the argument of the first exponential appearing in (\ref{FVsol}) is small for weak coupling. Employing a Taylor expansion it behaves as
\begin{eqnarray}
&& F_V(t,t;|\textbf{p}|) \, \simeq \, 
\nonumber\\
&& \frac{1}{2} \left\{ 1 - g^2 a\, e^{2\,\gamma_0\,t} \left[2\, {\rm daw}(\sqrt{2 \gamma_0 t}) - \frac{1}{\sqrt{2 \gamma_0 t}} \right]\right\} \, . \qquad
\label{eq:timesone} 
\end{eqnarray}
Accordingly, the fermion occupation number $n_{\psi}(t,|\textbf{p}|)$ given by (\ref{eq:occferm}) grows exponentially with approximately the {\em same rate} as the maximum primary growth rate for the bosons:
\begin{eqnarray}\nonumber
n_\psi(t,|\textbf{p}|) &\simeq& \frac{g^2 a}{2}\, e^{2\,\gamma_0\,t} \left[2\, {\rm daw}(\sqrt{2 \gamma_0 t}) - \frac{1}{\sqrt{2 \gamma_0 t}} \right] \\
&\simeq& \frac{g^2 a}{4 (2 \gamma_0 t)^{3/2}}e^{2\,\gamma_0\,t}\, .
\label{eq:npsione}
\end{eqnarray}
To arrive at the second approximate equality we used (\ref{eq:asymptotic}).
The exponential growth is parametrically slowly modified by the factor $\sim t^{-3/2}$, reducing somewhat the growth.
By fitting the time dependence to just an exponential in the respective regimes (see e.g.\ Fig.~\ref{fig:nf_analyt_diffg}) one may infer an approximate effective growth rate for the fermion modes of about $\gamma_{\psi}\approx 0.85 \gamma_0$, i.e.\ smaller than the maximum primary boson growth rate.
 
Because of the exponential growth the employed Taylor expansion that we used to obtain (\ref{eq:timesone}) or (\ref{eq:npsione}) becomes invalid for the regime (II). Without the Taylor expansion, using (\ref{eq:asymptotic}) in (\ref{FVsol}), one finds   
\begin{equation}
n_\psi(t,|\textbf{p}|) \simeq \frac{1}{2} - \frac{1}{2} \exp\left\{ - \frac{g^2 a}{2 (2 \gamma_0 t)^{3/2}}\, e^{2\,\gamma_0\,t}  \right\} .
\label{eq:doubleexp}
\end{equation}  
The asymptotic value $n_\psi(t,|\textbf{p}|) = 1/2$ is approached from below with a double-exponential time dependence. This corresponds to a very abrupt shutoff of the exponential growth period, as can also be seen in the behavior of the numerical solutions in Figs.~\ref{fig:nf_diffp} and \ref{fig:nf_analyt_diffp}. 

The characteristic time $t_\psi$ for the end of the exponential growth can be estimated to be the time when the argument of the first exponential in the solution (\ref{FVsol}) is of order one. Approximately this can be obtained from (\ref{eq:doubleexp}) by equating the absolute value of the argument of the first exponential to one, which yields
\begin{equation}\label{eq:t_psi}
t_\psi \, \simeq \, \frac{1}{2 \gamma_0} \ln\left( \frac{2}{g^2 a} \right) + \frac{3}{4 \gamma_0} \ln\left( 2\,\gamma_0\,t_\psi \right) \, .
\end{equation}
Though the growth rate does not depend on the choice of $c$ appearing in the dimensionless constant $a$ given by (\ref{eq:smalla}), we note that the time $t_\psi$ depends logarithmically on it. Using again $c \simeq 0.48$ we find $t_\psi \simeq 11.8$ for $g\sqrt{2}=0.001$, $t_\psi \simeq 9.2$ for $g\sqrt{2}=0.01$, and $t_\psi \simeq 6.7$ for $g\sqrt{2}=0.1$.

\begin{figure}[t]
  \includegraphics[scale=0.68]{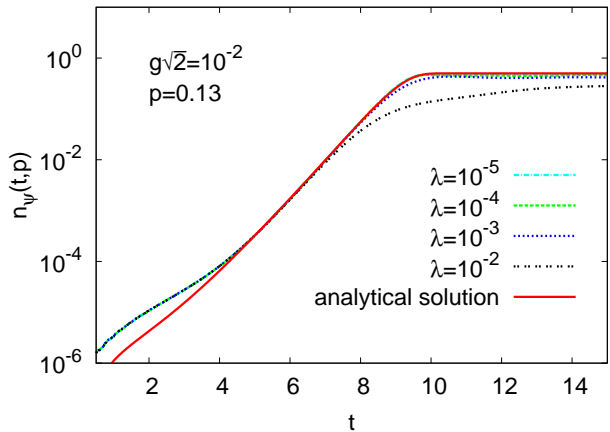}
\caption{Evolution of $n_{\psi}(t,|\textbf{p}|)$ as a function of time for given Yukawa coupling $g$ and different values for the boson self-coupling $\lambda$ as indicated in the graph. The solid line corresponds to the analytical solution and the dotted lines to the results from simulations.}\label{fig:nf_analyt_diffl}
\end{figure} 
In Fig.~\ref{fig:nf_analyt_diffg} we show a comparison of the analytical result (\ref{FVsol}) and the numerical simulation for different Yukawa couplings. The simulations are for the fermion mode with $|\textbf{p}|=0.13m$ (our smallest 
momentum) and a boson self-coupling of $\lambda=10^{-5}$.  One observes from Fig.~\ref{fig:nf_analyt_diffg} that the abrupt shutoff of the exponential growth is well reproduced for a wide range of Yukawa couplings $g$, and in these cases our above estimates for $t_\psi$ agree rather well with the numerical findings. However, sizeable deviations occur for $g\sqrt{2} < 10^{-3}$ with $\lambda = 10^{-5}$. Correspondingly, Fig.~\ref{fig:nf_analyt_diffl} shows results for fixed $g$ and different values of $\lambda$. The same statements apply as before, and similar deviations after the exponential growth terminates occur for $\lambda > 10^{-3}$ with $g\sqrt{2} = 10^{-2}$.
One observes that in both cases the ratio~\footnote{The ratio $g^2/\lambda$ is called resonance parameter in Ref.~\cite{Greene:1998nh}, since it plays also an important role for the resonant fermion production mechanism by coupling to a macroscopic scalar field, which is not discussed in our work.} 
\begin{equation}
\xi \, = \, \frac{g^2}{\lambda}
\label{eq:efficiency}
\end{equation}
at which significant corrections appear in the analytical description of the later stages of the fermion evolution is about $2\xi \simeq 0.1$. The limitation comes from the fact that we use the exponentially growing solution (\ref{eq:Fphismallp}) at all times, which is only correct before the growth of boson modes terminates, i.e.\ for times $t \lesssim t_\phi$ given by (\ref{eq:t_phi}). It is remarkable that according to Figs.~\ref{fig:nf_analyt_diffg} and \ref{fig:nf_analyt_diffl} this leads to practically no corrections to the solution (\ref{FVsol}) also for later times if $2\xi \gtrsim 0.1$. 
These findings can be understood from the fact that the characteristic time $t_\psi$ at which the growth of fermion occupation numbers terminates is earlier than $t_\phi$ in these cases. We use (\ref{eq:t_psi}) together with (\ref{eq:t_phi}) to estimate for which range of couplings one finds $t_\phi \gtrsim t_\psi$. Without further approximations this condition translates into the compact expression $\xi \gtrsim 1/(8 c)$, which is in reasonable agreement with the numerical findings as explained above.  

\begin{figure}[t]
 \includegraphics[scale=0.68]{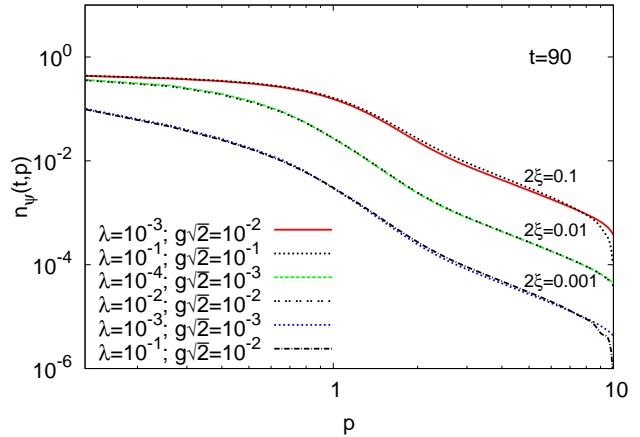}
 \caption{Fermion occupation numbers as a function of momentum at fixed time $t = 90/m$ after the exponential growth period. Compared are simulations
with different $\lambda$ and $g$ but equal $\xi = g^2/\lambda$.}\label{fig:nf_xi_comp}
\end{figure}
It is a general fact that the ratio (\ref{eq:efficiency}) plays an important role for the fermion dynamics for $t \gtrsim t_\phi$. This can be understood as follows. According to the discussion in Sec.~\ref{sec:bose_instability} at the time $t_\phi$ the parametric coupling-dependence of the boson correlation function becomes $F_\phi \sim {\cal O}(1/\lambda)$. The boson-fermion loop contribution to the fermion propagators shown in Fig.~\ref{fig:fermi_loop_eom} is proportional to $g^2$. Since for the dominant term the boson propagator line appearing in the loop is associated to $F_\phi$, this leads to the parametric $g^2/\lambda$ or $\xi$ dependence. In order to verify this we have plotted in Fig.~\ref{fig:nf_xi_comp} fermion occupation number distributions for different values of $g$ and $\lambda$. Compared are different simulations with same ratio $\xi$ for $2\xi = 0.1$, $0.01$ and $0.001$ at fixed time $t = 90/m$ after the exponential growth period. The agreement is remarkably good. We emphasize that at early times they differ in general for the same $\xi$. In particular, the time $t_\phi$ itself depends on $\lambda$ and is insensitive to $g$ for the considered weak couplings (see Sec.~\ref{sec:bose_instability}).      

\subsection{Emergence of power-law behavior}
\label{sec:fermi_pl}

We have seen in Sec.~\ref{sec:subsequent_pl} that boson occupation numbers approach a nonthermal scaling solution $\sim p^{-\kappa_{\text{IR}}}$ with occupation number exponent $\kappa_{\text{IR}} \simeq 4$ for low-momentum modes after the exponential growth period.
The corresponding nonthermal infrared fixed point has the dramatic consequence that a diverging time scale exists far from equilibrium, which can prevent or substantially delay thermalization due to critical slowing down~\cite{Berges:2008wm,Berges:2008sr}.
Because fermion field degrees of freedom obey the Pauli exclusion principle their occupation numbers cannot become large~\footnote{Of course, this does not concern fermions that can form bosonic bound states.}. This preempts infrared scaling solutions, which diverge in the infrared. As a consequence, the phenomenon of critical slowing down is expected to be absent for fermions. Low-momentum fermion modes may thus approach a thermal distribution much faster than bosons in this case, which we observed already in Sec.~\ref{sec:fermi_instability} and discuss in more detail below. 

In contrast to infrared scaling solutions, power-law behavior for higher momentum modes, where occupation numbers are not large, is of course not excluded for fermions. In the right column of Fig.~\ref{fig:spectra_comp}, which shows the evolution of $n_\psi(t,|{\textbf  p}|)$, the fermion occupation numbers exhibit a power-law regime with exponent two for momenta $|{\textbf  p}| \gtrsim m$. We have verified that the very sharp fall-off at later times of the distribution close to the cut-off, i.e. $|{\textbf  p}|\approx 10m$ in Fig.~\ref{fig:spectra_comp},   
is removed if we enlarge the momentum cutoff and the power-law behavior extends to larger momenta accordingly. One observes from Fig.~\ref{fig:spectra_comp} that the emergence of power-law behavior happens quite early, i.e. around the time $t\approx3/m$ and it is well established already around $t\approx5/m$.  This points out that the origin of the fermion power-law is distinct from the scaling behavior of the bosons discussed in Sec.~\ref{sec:subsequent_pl}.

In the following we will explain the fermion power-law regime. From Fig.~\ref{fig:nf_diffp}, in particular from the upper graph of that figure, one observes that the growth of higher momentum modes sets in later than for lower momenta. Since the growth rate is approximately momentum independent (see Sec.~\ref{sec:fermi_instability}) the retardation in the initial growth time translates into a momentum dependent amplitude 
\begin{equation}
A(|{\textbf  p}|) \, =\, \frac{n_\psi(t,|{\textbf p}|)}{e^{2 \gamma_0 t}} . 
\label{eq:ap}
\end{equation}
The residual time-dependence of this amplitude is parametrically slow compared to the dominant exponential growth behavior of $n_\psi(t,|{\textbf p}|)$. The observed power-law with exponent two then corresponds to $A(|{\textbf  p|}) \sim |{\textbf  p}|^{-2}$ in the respective momentum range with $|{\textbf  p}| \gtrsim m$. A further important property, which may be observed from Fig.~\ref{fig:nf_diffp} but better from the right column of Fig.~\ref{fig:spectra_comp}, is that the time for the end of the exponential growth period becomes approximately independent of momentum for not too small momenta. As a consequence, part of the momentum-dependence emerging at early times becomes "frozen-in" in time during the subsequent quasi-stationary evolution. In order to quantify these statements, we repeat the analytical calculation of Sec.~\ref{sec:fermi_instability} with the main difference that we take retardation or memory effects into account. To keep the discussion analytically tractable, our approximations will describe the dominant exponential time-dependence while subleading powers in time are discarded in contrast to what was done in Sec.~\ref{sec:fermi_instability}. This does not affect the relevant momentum dependence of $A(|{\textbf  p|})$ but overestimates the overall size of the amplitude. 

Starting point of the analysis is again the equation of motion for $F_V(t,t;|{\textbf  p}|)$ presented in (\ref{eq:Fphiappeq}). Approximating the 
boson two-point function $F_\phi(t,t';|{\textbf  p}|)$ by its primary-growth behavior (\ref{Fphi_linear_approx_sol}) and applying a memory expansion as in Sec.~\ref{sec:fermi_instability} would lead to the local differential equation with respect to momentum (\ref{FV_eq_final}), such that the momentum-independent solution (\ref{FVsol}) is obtained for the employed initial conditions. 
While this is a good approximation for the considered low-momentum behavior with $|{\textbf p}| \lesssim m$ in Sec.~\ref{sec:fermi_instability}, here we concentrate on larger momenta.
In order to make analytical progress, a perturbative estimate of the fermion correlators under the integral (\ref{eq:Fphiappeq}) is employed for small $g$. 
This amounts to replacing the respective correlators by their free-field solutions:
\begin{eqnarray}\label{free_fields_FV}
F_V(t,t'';|\textbf{p}|) & \rightarrow & \frac{1}{2}\cos\Bigl(|\textbf{p}|(t-t'')\Bigr)\, ,\\\label{free_fields_FV0}
F_V^0(t,t'';|\textbf{p}|) & \rightarrow & -\frac{i}{2} \sin\Bigl(|\textbf{p}|(t-t'')\Bigr)\, , \\\label{free_fields_rhoV}
\rho_V(t,t'';|\textbf{p}|) & \rightarrow & \sin\Bigl(|\textbf{p}|(t-t'')\Bigr)\, , \\\label{free_fields_rhoV0}
\rho_V^0(t,t'';|\textbf{p}|) & \rightarrow & i\cos\Bigl(|\textbf{p}|(t-t'')\Bigr)\, .
\end{eqnarray}
With these expressions the equation of motion (\ref{eq:Fphiappeq}) becomes
\begin{align}\nonumber
&\partial_t  F_V(t,t;|{\textbf  p}|) \simeq \\
&-2g^2\!\int_{0}^{t}dt''\!\int_{{\textbf q}} F_{\phi}(t,t'';|{\textbf p}-{\textbf q}|)\cos\Bigl(2|\textbf{p}|(t-t'')\Bigr)\, .\label{eq:FV_approx_large_p}
\end{align}
Using accordingly the lowest-order behavior (\ref{Fphi_linear_approx_sol}) for the scalar correlator, where $F_{\phi}(t,t'';|{\textbf p}| > m)$ vanishes or may be approximated by a Gaussian as in (\ref{def:K}), the momentum integral in (\ref{eq:FV_approx_large_p}) can be directly evaluated. However, the following time-integration would lead to expressions which are rather inconvenient to handle. Therefore, we may simplify the approximation by using $\gamma({\textbf p})\simeq \gamma_0$ in the momentum integral in (\ref{eq:FV_approx_large_p}), such that
\begin{equation}\label{p_int_approx2}
 \int_{{\textbf q}} F_{\phi}(t,t'';|{\textbf p}-{\textbf q}|) \, \simeq \, \frac{A_0 \gamma_0^3}{6 \pi^2} e^{\gamma_0(t+t'')}\, .
\end{equation}
Though this captures the dominant exponential time-dependence, comparison to (\ref{def:K}) shows that this neglects a parametrically slow $(t+t'')^{-3/2}$-correction. We have verified that this does not affect our main conclusions about momentum-dependencies, however, it overestimates the overall growth. Pursuing with (\ref{p_int_approx2}) in the equation of motion (\ref{eq:FV_approx_large_p}) the memory integral can be evaluated and gives the compact expression 
\begin{align}\nonumber
&\partial_t  F_V(t,t;|{\textbf p}|) \, \simeq \, -\frac{g^2A_0\gamma_0^3}{3\pi^2}\\
&\times\frac{\gamma_0e^{2\gamma_0\,t}\!-\!e^{\gamma_0\,t}\Bigl[ \gamma_0\cos\bigl(2|{\textbf p}|t\bigr)\!-\!2|\textbf{p}|\sin\bigl(2|{\textbf p}|t\bigr)\Bigr] }{4|{\textbf p}|^2+\gamma_0^2} \, . \label{eq:FVp}
\end{align}
Using the initial condition (\ref{FV_init}) this can be straightforwardly integrated and one obtains for the fermion occupation number (\ref{eq:occferm}):
\begin{align}\nonumber
n_\psi(t,|{\textbf p}|)& \, = \,
\frac{g^2A_0\gamma_0^3}{6\pi^2}\frac{e^{2\gamma_0\,t}-2e^{\gamma_0\,t}\cos\bigl(2|{\textbf p}|t\bigr)+1}{4|{\textbf p}|^2+\gamma_0^2}\\
&\, \simeq \, \frac{g^2A_0\gamma_0^3}{6\pi^2}\frac{e^{2\gamma_0\,t}}{4|{\textbf p}|^2+\gamma_0^2}\, .\label{nf_final}
\end{align}
While the first line in (\ref{nf_final}) represents a solution of (\ref{eq:FVp}) without further approximations, for the second line we exploit the fact that the dominant exponential term quickly takes over. Thus the oscillatory behavior becomes suppressed. 
We conclude from (\ref{nf_final}) that the amplitude (\ref{eq:ap})
\begin{equation}
A(|{\textbf  p}|) \, \sim \,  \frac{1}{4|{\textbf p}|^2+\gamma_0^2} \, , 
\label{eq:aest}
\end{equation}
which is in reasonable agreement with the behavior of the numerical solution displayed in the right column of Fig.~\ref{fig:spectra_comp}. In particular, with $\gamma_0 = m$ it describes the observed power-law exponent two for $|{\textbf p}|$ large enough compared to $m$. We emphasize that the above perturbative evaluation using (\ref{free_fields_FV})--(\ref{free_fields_rhoV0}) cannot describe the termination of the exponential growth period, in contrast to the self-consistent estimate of Sec.~\ref{sec:fermi_instability} leading to (\ref{FVsol}).   

\begin{figure}[t]
\begin{center}
\includegraphics[scale=0.68]{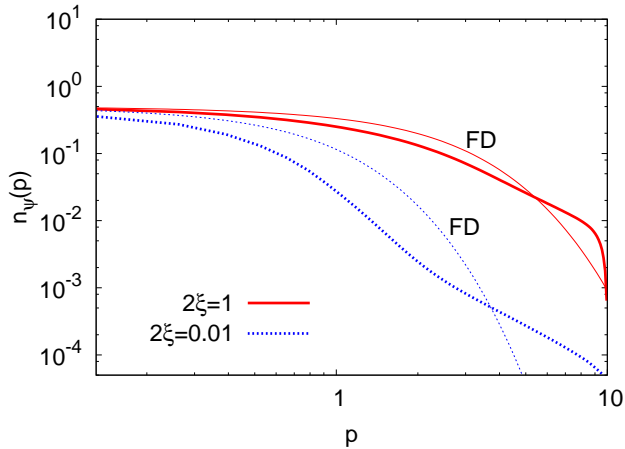}
\end{center}
\caption{Comparison between Fermi-Dirac distribution (thin) and simulation of the fermion occupation number at $t=90/m$ (thick) for $2\xi=1$ (solid line) and $2\xi=0.01$ (dotted line).}\label{fig:fermi_therm_comp}
\end{figure}
We have mentioned above that low-momentum fermion modes may approach a thermal distribution much faster than bosons. More precisely, the observed quasi-stationary solution then shares the property of the equilibrium solution that $n_\psi(|{\bf p}|=0)=1/2$ in the infrared \footnote{This is in contrast to typical oscillatory solutions oscillating between $n_\psi=0$ and $n_\psi=1$, as observed e.g. in \cite{Greene:1998nh} or \cite{GarciaBellido:2000dc}, in the absence of direct scattering. Whether the inclusion of scattering leads to a quasi-stationary evolution, where ensemble averages may be approximated by time averages with $n_\psi(|{\bf p}|=0)=1/2$ as in our work, is an interesting question.}. Following the discussion of Sec.~\ref{sec:fermi_instability} this is expected for couplings such that the ratio (\ref{eq:efficiency}) is $2\xi \gtrsim 0.1$. In this case the fermion occupation number growth terminates when the thermal equilibrium distribution is approached in the infrared. In order to quantify this aspect we compare in Fig.~\ref{fig:fermi_therm_comp} a Fermi-Dirac distribution with the occupation number $n_\psi (t,|{\bf p}|)$ in the quasi-stationary regime at $t = 90/m$. The results for $2\xi=1$ are obtained from $g\sqrt{2}=0.01$ and $\lambda=10^{-4}$, whereas $2\xi=0.01$ results from $g\sqrt{2}=0.01$ and $\lambda=0.01$. More precisely, we compare to the thermal distribution of free massless particles with zero net charge, 
\begin{eqnarray}\label{n_FD}
 n_{\text{FD}}(|\textbf{p}|)&=&\frac{1}{e^{|\textbf{p}|/T}+1}
\,\, \stackrel{|\textbf{p}|\rightarrow 0}{\longrightarrow}\,\, \frac{1}{2}\, ,
\end{eqnarray}
where the temperature is estimated by equating the respective energies obtained in a quasi-particle picture,
\begin{equation}\label{eq:temp_det}
 \int\frac{d^3p}{(2\pi)^3}\,|\textbf{p}|\,n_{\text{FD}}(|\textbf{p}|)\,= \, \int\frac{d^3p}{(2\pi)^3}\,|\textbf{p}|\,n_{\psi}(t,|\textbf{p}|).
\end{equation}
Fig.~\ref{fig:fermi_therm_comp} indeed shows that for $2\xi = 1$ the quasi-stationary occupation number follows the thermal distribution rather closely up to momenta $|{\textbf p}| \simeq m$. This has to be compared to the behavior of the bosons, which still evolve in the infrared orders of magnitude in time longer towards the nonthermal fixed point distribution $n_\phi(t,|{\textbf p}|) \sim |{\textbf p}|^{-4}$, which is far from thermal equilibrium (see Sec.~\ref{sec:subsequent_pl}). Fig.~\ref{fig:fermi_therm_comp} shows that for $2\xi = 0.01$ the exponential fermion occupation number growth stops before $n_{\rm FD}(|{\textbf p}|=0) = 1/2$ is reached. Accordingly, the deviations from the thermal distribution are larger in the infrared and also the differences at higher momenta due to the nonthermal power-law behavior are even more pronounced. Phenomenologically a major aspect is that the power-law leads to a substantial excess of high-momentum particles compared to thermally produced particles.

\section{Conclusions}
\label{sec:conclusions}

We have studied nonequilibrium fermion production in a Yukawa-type model in 
$3+1$ dimensions following a
spinodal/tachyonic instability. The approximation is based on a $1/N$ expansion to NLO of 
the 2PI effective action
out of equilibrium, including the two-loop boson-fermion fluctuations displayed in Fig.~\ref{fig:feyn_diag}. This resums an infinite series of scattering processes and takes into account memory effects. 

These quantum corrections turned out to be crucial for the phenomenon of instability-induced fermion production observed in this work. This production mechanism is based on the exponential growth of long-wavelength boson occupation numbers following a nonequilibrium instability. A major characteristic property is that the amplification of fermion occupation numbers is approximately momentum-independent and proceeds with the maximum primary growth rate of boson modes. 

Depending on the ratio $\xi$ of the square of the Yukawa coupling and the boson self-coupling, the end of the exponential amplification of fermion modes is either determined by the end of the 
boson growth period ($2\xi \lesssim 0.1$), or by reaching the Fermi-Dirac distribution in the infrared for momenta $|{\textbf p}| \lesssim m$ ($2\xi \gtrsim 0.1$). The latter leads to very fast thermalization of low-momentum modes. This finding has to be confronted with the extremely slow thermalization of low-momentum bosons, which are still far from equilibrium at this time. 

The slowing-down of the boson dynamics after the exponential growth period is shown to be associated to the approach to a nonthermal infrared fixed point. It is characterized by scaling of occupation number modes $n_\phi(|{\textbf p}|) \sim |{\textbf p}|^{-4}$, which lead to strongly enhanced fluctuations in the infrared as compared to a thermal low-momentum distribution $\sim |{\textbf p}|^{-1}$. It was pointed out in Ref.~\cite{Berges:2008wm} that such a fixed point is approached in scalar models after a parametric resonance instability, which can prevent or substantially delay thermalization. Here we have demonstrated that a spinodal/tachyonic instability leads to scaling with the same occupation number exponent, which suggests that this is a universal phenomenon of nonequilibrium instability dynamics. We note that for the spinodal/tachyonic instability and the considered range of parameters we seem not to observe the expected phenomenon of perturbative Kolmogorov wave turbulence at higher momenta~\cite{Micha:2002ey}. Instead boson modes with momenta $|{\textbf p}| \gtrsim m$ exhibit thermal properties at this stage. 

Fermion occupation number modes are bounded from above, which preempts infrared scaling solutions. We have explicitly verified that the absence of a scaling regime for fermions at low momenta does not affect the infrared scaling behavior observed for bosons. In general we find that the boson dynamics is rather insensitive to the presence of fermions for not too large Yukawa coupling $g\lesssim 1$, which is expected because of the comparably large boson occupation numbers. Despite the absence of an infrared scaling regime for fermions, we find a power-law behavior with exponent two for higher momenta $|{\textbf p}| \gtrsim m$. Its origin is very distinct from the physics underlying nonthermal fixed points. The latter appear at late times and can be described as stationary points of time evolution equations that are independent of the initial conditions. In contrast, we could show that the observed fermion power-law behavior emerges at very early times and is a consequence of a retardation or memory effect. This power-law leads to an excess of highly energetic particles compared to thermal distributions. 

It is remarkable that for the considered model these nonlinear out-of-equilibrium phenomena  can be studied numerically -- and to a large extent analytically -- using 2PI effective action techniques directly in quantum field theory. An important improvement would be to include a macroscopic field into our analysis, in order to estimate the importance of quantum corrections to frequently employed approximations based on the Dirac equation in the presence of classical fields. Understanding the impact of quantum fluctuations is an important step towards more realistic phenomenological applications in early universe cosmology as well as QCD in the context of collision experiments of heavy nuclei.\\

We thank Szabolcs Borsanyi, Sebastian Scheffler, Denes Sexty and Thorsten Z{\"o}ller for many helpful discussions and collaboration on related work. This work is supported in part by the Yukawa International Program for Quark-Hadron Sciences at the Yukawa Institute for Theoretical Physics, Kyoto University, by the BMBF grant 06DA267, and by the DFG under contract SFB634.
Some of the ideas were initiated during the program
on "Nonequilibrium Dynamics in Particle Physics and
Cosmology" (2008) at the Kavli Institute for Theoretical Physics in Santa Barbara, 
supported by the NSF under grant PHY05-51164.

\appendix

\section{Time evolution of the spectral function}
\label{sec:Fphi_dom}
\begin{figure}[t]
\begin{center}
 \includegraphics[scale=0.68]{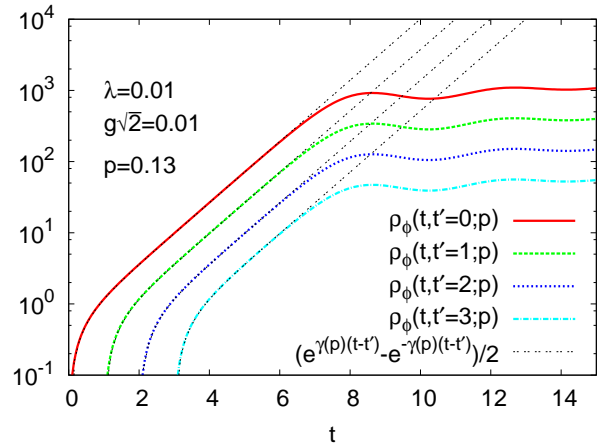}
\end{center}
\caption{Time evolution of the unequal-time correlation function $\rho_\phi(t,t';|{\bf p}|)$ for different $t'$ with couplings and momentum as indicated in the figure. For each $t'$ the analytical expression (\ref{Rho_linear_approx}) is plotted for comparison.}\label{fig:Rhophi_linear_analyt_comp}
\end{figure}
\begin{figure}[t]
\begin{center}
 \includegraphics[scale=0.68]{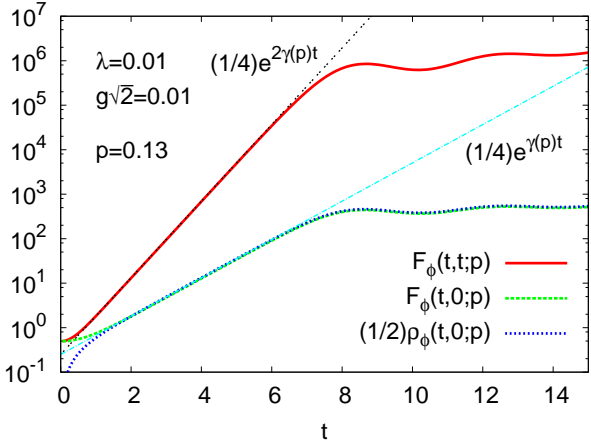}
\end{center}
\caption{Comparison of the time evolution of the statistical and spectral function together with the analytical expressions
discussed in the text. Couplings and the chosen momentum are indicated in the figure.}\label{fig:Fphi_Rhophi_linear_comp}
\end{figure}

 
In this Appendix we consider the evolution of the unequal-time boson spectral function  
$\rho_\phi(t,t';|{\bf p}|)$ and statistical function $F_\phi(t,t';|{\bf p}|)$. We will use this to explain in more detail some estimates in the main text based on the dominance of contributions coming from the equal-time correlator $F_\phi(t,t;|{\bf p}|)$.
According to (\ref{sym_rel_b}) the 
equal-time spectral function $\rho_\phi(t,t;|{\bf p}|)$ vanishes identically.
The time evolution of $F_\phi(t,t';|{\bf p}|)$ in the linear regime can be approximated
by (\ref{Fphi_linear_approx_sol}). Similar considerations for the spectral function lead to 
\begin{equation}\label{Rho_linear_approx}
 \rho_\phi(t,t';|{\bf p}|)= \frac{1}{2}\left(e^{\gamma({\bf p})(t-t')}-e^{-\gamma({\bf p})(t-t')} \right) \, ,
\end{equation}
with $\gamma({\bf p})$ defined in (\ref{eq:growthrate}).
In Fig.~\ref{fig:Rhophi_linear_analyt_comp} we plot results from simulations together with the corresponding analytical expression (\ref{Rho_linear_approx}) for
couplings and momentum as indicated in the figure. The agreement before the exponential amplification terminates is remarkable for low-momentum modes. 
For momenta $|{\bf p}|\gtrsim m$ the approximation becomes worse. 
With (\ref{Rho_linear_approx}) and (\ref{Fphi_linear_approx_sol})
for the statistical function one observes that for not too early times $F_\phi(t,t;|{\bf p}|)$ dominates over
$F_\phi(t,t';|{\bf p}|)$ and $\rho_\phi(t,t';|{\bf p}|)$. Moreover one can find in this time regime the relation
\begin{equation}\label{eq:Fphit0_Rhophit0}
 F_\phi(t,0;|{\bf p}|) \simeq \frac{1}{2}\rho_\phi(t,0;|{\bf p}|)\, ,
\end{equation}
which holds also for large times. In Fig.~\ref{fig:Fphi_Rhophi_linear_comp} 
the correlation functions $F_\phi(t,t;|{\bf p}|)$, $F_\phi(t,0;|{\bf p}|)$ and $\rho_\phi(t,0;|{\bf p}|)$ are
plotted together with (\ref{Fphi_linear_approx_sol}) and the dominant part of (\ref{Rho_linear_approx}) with a factor $1/2$. This confirms (\ref{eq:Fphit0_Rhophit0}) and the dominance of $F_\phi(t,t;|{\bf p}|)$ is well established for not too early times.
We note that for $t'>0$ the correlator $F_\phi(t,t';|{\bf p}|)$ becomes larger than $F_\phi(t,0;|{\bf p}|)$, whereas $\rho_\phi(t,t';|{\bf p}|)$
becomes smaller than $\rho_\phi(t,0;|{\bf p}|)$ as Fig.~\ref{fig:Rhophi_linear_analyt_comp} shows. These findings justify the neglection of terms proportional to $\rho_\phi(t,t';|{\bf p}|)$ in the self-energy contributions in (\ref{FV_eom}), as was done for the analytical estimates in Sec.~\ref{sec:fermion_dynamics}.

\section{Decomposition of fermion two-point functions}
\label{sec:LR_decomp}
In this Appendix we present the expressions of the statistical- and spectral-function of the fermion two-point function
in the chiral basis. 
The statistical function is defined by (we omit Dirac indices)
\begin{equation}
 F(x,y)=\frac{1}{2}\bigg\langle \left[ \psi(x),\bar{\psi}(y)\right] \bigg\rangle,
\end{equation}
where the expectation value has to be taken with respect to the initial density matrix and $\bar{\psi}(x)\equiv\psi^{\dagger}(x)\gamma_0$.
In the chiral basis we can write $\psi(x)=\psi_{L}(x)+\psi_{R}(x)$. Where
\begin{equation}
\psi_{L}(x)=\begin{pmatrix}
\chi_{L}(x) \\ 0 
\end{pmatrix} \quad \text{and}\quad
\psi_{R}(x)=\begin{pmatrix}
0 \\ \chi_{R}(x) 
\end{pmatrix}.
\end{equation}
The $\chi_{L/R}(x)$ are two-component Weyl spinors. 
This decomposition is done by projections ($P_L$ and $P_R$) given as
\begin{eqnarray}\nonumber
 \psi(x)&=&P_L\psi(x)+P_R\psi(x)\\
&\equiv&\frac{1}{2}\left(1-\gamma_5 \right)\psi(x)+\frac{1}{2}\left(1+\gamma_5 \right)\psi(x).
\end{eqnarray}
The $\gamma$-matrices in this basis are given by
\begin{eqnarray}
 \gamma_0&=&\begin{pmatrix} 0 & \mathds{1}_{2\times2}  \\ \mathds{1}_{2\times2} & 0 \end{pmatrix}, \quad
\gamma_i=\begin{pmatrix} 0 & \sigma_i \\ -\sigma_i & 0\end{pmatrix}, \\\nonumber
\gamma_5&=&\begin{pmatrix} -\mathds{1}_{2\times2} & 0\\ 0 & \mathds{1}_{2\times2} \end{pmatrix}. \quad
\end{eqnarray}
Using this notation we get an expression for the statistical function in terms of left and right handed spinors.
Here we merely give the result of the vector components relevant in the analysis given in the main text.
The expression for the temporal component is 
\begin{eqnarray}\nonumber
&& F_V^0(x,y)=
\frac{1}{8}\left[\text{tr}\left(\bigg\langle\left[ \chi_{R}(x),\chi^{\dagger}_{R}(y)\right] \bigg\rangle\right)\right.\\
&&\left.+\text{tr}\left(\bigg\langle\left[\chi_{L}(x),\chi^{\dagger}_{L}(y)\right]\bigg \rangle\right)\right] ,
\end{eqnarray}
and for the spatial components one finds
\begin{eqnarray}\nonumber
&& \textbf{F}_V(x,y)=\frac{1}{8}\left[\text{tr}\left(\boldsymbol{\sigma}\bigg\langle\left\lbrace \chi_{R}(x),\chi^{\dagger}_{R}(y)\right\rbrace \bigg\rangle\right)\right.\\
&&\left.-\text{tr}\left(\boldsymbol{\sigma}\bigg\langle\left\lbrace \chi_{L}(x),\chi^{\dagger}_{L}(y)\right\rbrace \bigg\rangle\right)\right] .
\end{eqnarray}
The spectral function is defined by
\begin{equation}
 \rho(x,y)=i\bigg\langle \left\lbrace \psi(x),\bar{\psi}(y)\right\rbrace \bigg\rangle.
\end{equation}
Similarly as above, the temporal component of the spectral function is
\begin{eqnarray}\nonumber
&& \rho_V^0(x,y)=
i\left[\text{tr}\left(\bigg\langle\left\lbrace  \chi_{R}(x),\chi^{\dagger}_{R}(y)\right\rbrace  \bigg\rangle\right)\right.\\
&&\left.+\text{tr}\left(\bigg\langle\left\lbrace \chi_{L}(x),\chi^{\dagger}_{L}(y)\right\rbrace \bigg \rangle\right)\right] ,
\end{eqnarray}
and the spatial components are given by
\begin{eqnarray}\nonumber
&& \boldsymbol{\rho}_V(x,y)=i\left[\text{tr}\left(\boldsymbol{\sigma}\bigg\langle\left[ \chi_{R}(x),\chi^{\dagger}_{R}(y)\right] \bigg\rangle\right)\right.\\
&&\left.-\text{tr}\left(\boldsymbol{\sigma}\bigg\langle\left[ \chi_{L}(x),\chi^{\dagger}_{L}(y)\right] \bigg\rangle\right)\right] .
\end{eqnarray}

\section{Numerical method}
\label{sec:numerical}
Our numerical calculations for the equations of motion derived from the 2PI effective action are carried out on a single personal computer with 8Gb RAM. The discretization is performed on the level of the equations of motion. We discretize two time dimensions and use spherical symmetry for the three-dimensional momentum integrals, so that only the absolute value of the discretized momenta needs to remain in memory. The angular part of the integrals can be done analytically in each case because the arising convolutions in momentum space are given as a product of two functions in coordinate space. The transition between coordinate and momentum space is implemented using the FFTW library (http://www.fftw.org/) for one-dimensional Fourier transforms.
For the time dimension we use up to 1850 time steps with a width of $dt\leq 0.05$,
depending on the coupling constants. The momentum grid has $N=80$ discretization points with an equidistant
spacing of $dp=\pi/(N\,a_s)$, where $a_s=0.3$ is the spatial grid spacing.
We use a leap-frog-type algorithm following earlier work, for details see~\cite{Berges:2002wr,Berges:2004ce}. With this algorithm the fermion sector has a discretization in time with a time-step width of twice the one in the boson sector. All of the derivatives and integrals are calculated with the
Euler method and the trapezoid rule, respectively.

In the case of the classical-statistical simulations for bosonic fields (see Figs.~\ref{fig:2PI_vs_class} and \ref{fig:non-them-fixed}),
we use the standard hypercubic lattice in
3+1 dimensions. The employed Laplace operator in the classical field equation of motion is of fourth order and combined with a finite-difference second-order time derivative. The field initial conditions are generated to match the Gaussian initial correlators for the quantum 2PI NLO calculation
and a 'quench' is implemented by changing the sign of the boson mass within the two time-steps necessary to set up a second order time evolution. Averages for two-point functions are obtained by summing over all possible redundant combinations of distances or momenta, respectively. The necessary
transition between coordinate
and momentum space is implemented using the FFTW library for three-dimensional Fourier transforms.
Our maximum possible lattice size on a 8Gb memory machine is 460 points along each spatial axis with the corresponding spacing $a_s=0.3$ whereas the each temporal step corresponds to $dt=0.05$.


\begin{thebibliography}{10}

\bibitem{Traschen:1990sw}
  J.~H.~Traschen and R.~H.~Brandenberger,
  Phys.\ Rev.\  D {\bf 42} (1990) 2491.

\bibitem{Kofman:1994rk}
  L.~Kofman, A.~D.~Linde and A.~A.~Starobinsky,
  Phys.\ Rev.\ Lett.\  {\bf 73} (1994) 3195.

\bibitem{Boyanovsky:1994me}
  D.~Boyanovsky, H.~J.~de Vega, R.~Holman, D.~S.~Lee and A.~Singh,
  Phys.\ Rev.\  D {\bf 51} (1995) 4419.

\bibitem{Khlebnikov:1996mc}
  S.~Y.~Khlebnikov and I.~I.~Tkachev,
  Phys.\ Rev.\ Lett.\  {\bf 77} (1996) 219.
     
\bibitem{Berges:2002cz}
  J.~Berges and J.~Serreau,
  Phys.\ Rev.\ Lett.\  {\bf 91} (2003) 111601.

\bibitem{Plasmainst}
S.~Mr\'owczy\'nski, Phys.\ Lett.\ B {\bf 214} (1988) 587; Y.~E.~Pokrovsky and A.~V.~Selikhov, JETP Lett. 47, 12 (1988) [Pisma Zh. Eksp. Teor. Fiz. 47, 11 (1988)].

\bibitem{Plasmainst2}
P.~Arnold, J.~Lenaghan and G.~D.~Moore, JHEP {\bf 08} (2003) 002;
P.~Romatschke and M.~Strickland, Phys.\ Rev.\ D {\bf 68} (2003) 036004.

\bibitem{WongYangMills}
A.~Dumitru and Y.~Nara, Phys.\ Lett.\ B {\bf 621} (2005) 89;
A.~Dumitru, Y.~Nara and M.~Strickland, Phys.\ Rev.\ D {\bf 75} (2007) 025016.

\bibitem{Romatschke:2006nk}
  P.~Romatschke and R.~Venugopalan,
  Phys.\ Rev.\ Lett.\ {\bf 96} (2006) 062302;
  {\em ibid.}, Phys.\ Rev.\ D {\bf 74} (2006) 045011.

\bibitem{Berges:2007re}
  J.~Berges, S.~Scheffler and D.~Sexty,
  Phys.\ Rev.\  D {\bf 77} (2008) 034504;
  J.~Berges, D.~Gelfand, S.~Scheffler and D.~Sexty,
  arXiv:0812.3859 [hep-ph].

\bibitem{Fujii:2008dd}
  H.~Fujii and K.~Itakura,
  Nucl.\ Phys.\  A {\bf 809} (2008) 88;
  H.~Fujii, K.~Itakura and A.~Iwazaki,
  arXiv:0903.2930 [hep-ph].

\bibitem{Micha:2002ey}
  R.~Micha and I.~I.~Tkachev,
  Phys.\ Rev.\ Lett.\  {\bf 90} (2003) 121301;
  {\it ibid.} Phys.\ Rev.\  D {\bf 70} (2004) 043538.

\bibitem{Berges:2008wm}
  J.~Berges, A.~Rothkopf and J.~Schmidt,
  Phys.\ Rev.\ Lett.\  {\bf 101} (2008) 041603.

\bibitem{Barnaby:2009mc}
  N.~Barnaby, Z.~Huang, L.~Kofman and D.~Pogosyan,
  arXiv:0902.0615 [hep-th].

\bibitem{Arnold:2005qs}
P.~Arnold, G.~D.~Moore, Phys.\ Rev.\ D {\bf 73} (2006) 025006;

\bibitem{Berges:2008mr}
  J.~Berges, S.~Scheffler and D.~Sexty,
  arXiv:0811.4293 [hep-ph].

\bibitem{Berges:2008sr}
  J.~Berges and G.~Hoffmeister,
  Nucl.\ Phys.\  B {\bf 813} (2009) 383. 
    
\bibitem{Berges:2001fi}
  J.~Berges,
  Nucl.\ Phys.\  A {\bf 699} (2002) 847.

\bibitem{Berges:2002wr}
  J.~Berges, S.~Borsanyi and J.~Serreau,
  Nucl.\ Phys.\  B {\bf 660} (2003) 51.

\bibitem{Mrowczynski:2001az}
  S.~Mrowczynski,
  Phys.\ Rev.\  D {\bf 65} (2002) 117501.

\bibitem{Schenke:2006fz}
  B.~Schenke and M.~Strickland,
  Phys.\ Rev.\  D {\bf 74} (2006) 065004.


\bibitem{Boyanovsky:1995ema}
  D.~Boyanovsky, M.~D'Attanasio, H.~J.~de Vega, R.~Holman and D.~S.~Lee,
  Phys.\ Rev.\  D {\bf 52} (1995) 6805.

\bibitem{Baacke:1998di}
  J.~Baacke, K.~Heitmann and C.~Patzold,
  Phys.\ Rev.\  D {\bf 58} (1998) 125013.
  
\bibitem{Greene:1998nh}
  P.~B.~Greene and L.~Kofman,
  Phys.\ Lett.\  B {\bf 448} (1999) 6;
  {\em ibid.} Phys.\ Rev.\  D {\bf 62} (2000) 123516.
  
\bibitem{Giudice:1999fb}
  G.~F.~Giudice, M.~Peloso, A.~Riotto and I.~Tkachev,
  JHEP {\bf 9908} (1999) 014; 
  G.~F.~Giudice, A.~Riotto and I.~Tkachev,
  JHEP {\bf 9911} (1999) 036.

\bibitem{GarciaBellido:2000dc}
  J.~Garcia-Bellido, S.~Mollerach and E.~Roulet,
  JHEP {\bf 0002} (2000) 034. 

\bibitem{Peloso:2000hy}
  M.~Peloso and L.~Sorbo,
  JHEP {\bf 0005} (2000) 016. 

\bibitem{Tsujikawa:2000ik}
  S.~Tsujikawa, B.~A.~Bassett and F.~Viniegra,
  JHEP {\bf 0008} (2000) 019. 

\bibitem{BasteroGil:2000je}
  M.~Bastero-Gil and A.~Mazumdar,
  Phys.\ Rev.\  D {\bf 62} (2000) 083510. 

\bibitem{Nilles:2001ry}
  H.~P.~Nilles, M.~Peloso and L.~Sorbo,
  Phys.\ Rev.\ Lett.\  {\bf 87} (2001) 051302; 
  {\em ibid.} JHEP {\bf 0104} (2001) 004.

\bibitem{Baacke:2007ca}
  J.~Baacke, N.~Kevlishvili and J.~Pruschke,
  JCAP {\bf 0706} (2007) 004. 

\bibitem{Gelis:2005pb}
  F.~Gelis, K.~Kajantie and T.~Lappi,
  Phys.\ Rev.\ Lett.\  {\bf 96} (2006) 032304.

\bibitem{Aarts:1998td}
  G.~Aarts and J.~Smit,
  Nucl.\ Phys.\  B {\bf 555} (1999) 355;
  {\em ibid.} Phys.\ Rev.\  D {\bf 61} (1999) 025002.

\bibitem{Ramsey:1997fz}
  S.~A.~Ramsey, B.~L.~Hu and A.~M.~Stylianopoulos,
  Phys.\ Rev.\  D {\bf 57} (1998) 6003.

\bibitem{Berges:2004ce}
  J.~Berges, S.~Borsanyi and C.~Wetterich,
  Phys.\ Rev.\ Lett.\  {\bf 93} (2004) 142002. 

\bibitem{Lindner:2007am}
  M.~Lindner and M.~M.~Muller,
  Phys.\ Rev.\  D {\bf 77} (2008) 025027.

\bibitem{Arrizabalaga:2004iw}
  A.~Arrizabalaga, J.~Smit and A.~Tranberg,
  JHEP {\bf 0410} (2004) 017.

\bibitem{Cornwall:1974vz}
  J.~M.~Cornwall, R.~Jackiw and E.~Tomboulis,
  Phys.\ Rev.\  D {\bf 10} (1974) 2428.

\bibitem{Luttinger:1960ua}
  J.~M.~Luttinger and J.~C.~Ward,
  Phys.\ Rev.\  {\bf 118} (1960) 1417;
  G.~Baym,
  Phys.\ Rev.\  {\bf 127} (1962) 1391.

\bibitem{Chou:1984es}
  K.~c.~Chou, Z.~b.~Su, B.~l.~Hao and L.~Yu,
  Phys.\ Rept.\  {\bf 118} (1985) 1.

\bibitem{Calzetta:1986cq}
  E.~Calzetta and B.~L.~Hu,
  Phys.\ Rev.\  D {\bf 37} (1988) 2878.

\bibitem{Berges:2004yj}
  J.~Berges,
  AIP Conf.\ Proc.\  {\bf 739} (2004) 3; arXiv:hep-ph/0409233.

\bibitem{Schwinger:1960qe}
  J.~S.~Schwinger,
  J.\ Math.\ Phys.\  {\bf 2} (1961) 407.
  L.~V.~Keldysh,
  Zh.\ Eksp.\ Teor.\ Fiz.\  {\bf 47} (1964) 1515 
  [Sov.\ Phys.\ JETP {\bf 20} (1965) 1018].

\bibitem{Aarts:2001qa}
  G.~Aarts and J.~Berges,
  Phys.\ Rev.\  D {\bf 64} (2001) 105010.

\bibitem{Aarts:2001yn} 
  G.~Aarts, J.~Berges,
  Phys.\ Rev.\ Lett.\  {\bf 88} (2002) 041603.

\bibitem{Berges:2007ym}
  J.~Berges and T.~Gasenzer,
  Phys.\ Rev.\  A {\bf 76} (2007) 033604.

\bibitem{Berges:1997eu}
  J.~Berges, D.~U.~Jungnickel and C.~Wetterich,
  Phys.\ Rev.\  D {\bf 59} (1999) 034010. 

\bibitem{Abramowitz:1964}
 M.~Abramowitz, I.~A.~Stegun,
Handbook of Mathematical Functions,
Dover, New York, 1972, 9th printing, pp.\ 295 and 319.


\end{thebibliography}
\end{document}